\documentclass[ twocolumn, twocolappendix]{aastex631}

\usepackage{amsmath}

\DeclareRobustCommand{\Mpc}{\mathrm{Mpc}}
\newcommand{\msun}{\hbox{$\mathrm{M}_{\odot}$}}
\newcommand{\lsun}{\hbox{$\mathrm{L}_{\odot}$}}

\DeclareRobustCommand{\kpc}{\mathrm{kpc}}

\DeclareRobustCommand{\gcmasshalomass}{$M_\mathrm{GC}$--$M_\mathrm{h}$}
\DeclareRobustCommand{\gcnumberhalomass}{$N_\mathrm{GC}$--$M_\mathrm{h}$}

\defcitealias{Harris_2017}{H17}

\begin{document}

\title{Globular Cluster counts around 700 Nearby Galaxies}

\author[0009-0003-0674-9813]{Minh Ngoc Le}
\affiliation{Institute of Astronomy,
National Tsing Hua University, Hsinchu 30013, Taiwan}
\email{lmngoc1509@gmail.com}

\author[0000-0001-8274-158X]{Andrew P. Cooper}
\affiliation{Institute of Astronomy,
National Tsing Hua University, Hsinchu 30013, Taiwan}
\affiliation{Department of Physics,
National Tsing Hua University, Hsinchu 30013, Taiwan}
\affiliation{Center for Informatics and Computation in Astronomy,
National Tsing Hua University, Hsinchu 30013, Taiwan}
\email{apcooper@gapp.nthu.edu.tw}



\begin{abstract}


Empirically, the total number (or total mass) of globular clusters bound in a single galactic system correlates with the virial mass of the system. The form of this relation and its intrinsic scatter are potentially valuable constraints on theories of globular cluster formation and galaxy evolution. In this work, we use the DESI Legacy Imaging Survey to make a large-scale, homogeneous estimate of GC abundance around 707 galaxies at distances $\lesssim 30\,\mathrm{Mpc}$ with luminosities $8 \leq \log_{10}L/\mathrm{L}_\odot \leq 11.5$. The combination of depth and sky coverage in DESI-LS allow us to extend the techniques used by previous ground-based photometric GC surveys to a larger and potentially more representative sample of galaxies.
We find average GC counts and radial profiles that are broadly consistent with the literature on individual galaxies, including good agreement with the distribution of GCs in the Milky Way, demonstrating the viability of DESI-LS images for this purpose. We find a relation between GC counts and virial mass in agreement with previous estimates based on heterogenous datasets, except at the lowest masses we probe, where we find a larger scatter in the number of cluster candidates and a slightly higher average count. 
\end{abstract}



\section{Introduction} 
\label{sec:intro}

Globular clusters (hereafter GCs) are self-bound objects comprising thousands to millions of stars of near-identical age confined within a characteristic radius of a few parsecs. More than 150 GCs are known in the Milky Way, many on weakly-bound, eccentric orbits resembling those of stars in the Galactic stellar halo \citep{Harris_MW_GCs, Baumgardt2021_GC_MW_profile}. GC systems of similar number and extent appear to be typical in galaxies comparable to the Milky Way \citep{Spitler:2009aa, Harris2013}. Some properties of GC systems appear remarkably uniform from one galaxy to the next, most notably the GC luminosity function (GCLF). Other properties, such as the distribution of GC colors and radial distributions of GCs around their host galaxies, are known to vary systematically with the virial mass of the host galaxy.
  
The formation of GCs requires a massive reservoir of dense gas to support an extremely rapid but short-lived burst of star formation. Such conditions are rare in present-day galaxies, but may have been more common at earlier epochs, consistent with the typically old ages of GCs inferred from their stellar populations. Beyond this very general constraint, little is known about where, when and how GCs formed. The prevailing view in the recent literature is that the majority form around `cosmic noon' ($1\lesssim z \lesssim 4$) in the progenitors of present-day massive galaxies, when patches of the interstellar medium (ISM) are driven to high density by turbulence and/or `shocks' that arise in galaxy mergers \citep{ Ashman:1992aa, Elmergreen1997ApJ...480..235E, Kavtsov_Gnedin2005, Muratov:2010aa, Kruijssen2015MNRAS}. Systems like the Milky Way are therefore expected (but have not yet been proven) to form a significant fraction of GCs `in situ'. Conversely, the GC systems of massive elliptical galaxies have likely been assembled only recently through a succession of gas-poor mergers between many less massive systems, much like the field stars in the same galaxies \citep[e.g.][]{De-Lucia:2007aa}. Differences in the star formation histories of those progenitors and those of the Milky Way may explain differences in the properties of such GC systems, over and above their greater number of clusters. 

A better understanding of GC formation may therefore help to constrain the nature of the star-forming ISM at high redshift \citep{Krumholz2019_GC_starformation}. Moreover, the fact that the stars in diffuse galactic stellar halos have similar ages, metal enrichment and characteristic orbits to GCs \citep[e.g.][]{Lynden-Bell:1995aa} suggests that a significant fraction of GCs have been accreted from less massive systems. At lower halo masses, the accreted GC population may include a signficant contribution froms `primoridal' population of nuclear star clusters \citep[e.g.][]{Zinnecker:1988aa,Kruijssen_Cooper_2012_initial_mass_GCs}. Since GCs and halo stars may be weighted towards different types or ages of progenitor galaxies, they could serve as mutually complementary tracers of the hierarchical assembly of DM halos \citep[e.g][]{Prieto:2008aa, Krujssen_trace_gal_by_GC, Kruijssen2020,Chen:2024aa}.

The subject of this paper is the apparently fundamental relation between the number (or combined mass) of GCs in a host galaxy and its virial mass\footnote{Or less precisely, dark matter halo mass.} \citep{Spitler:2009aa}. Hereafter we refer to this as the \gcnumberhalomass{} (or \gcmasshalomass{}) relation. Evidence for such a relation was recognized in early work on the GC populations of cluster galaxies  \citep[e.g.][]{West:1995aa,Blakeslee:1997aa,Blakeslee:1999aa,Peng:2008aa} and explored across a wider mass range by \citet{Spitler:2009aa}. More recently, \citet{Harris2013} collated observations of the GC systems of more than $400$ galaxies in the (approximate) virial mass range $5 \times 10^{10}$ to $ 10^{15} \ \msun{}$. A series of papers based on that dataset, concluding with \citet[][hearafter H17]{Harris_2017}, provide strong support for an approximately linear \gcmasshalomass{} relation\footnote{$M_\mathrm{GC}$ is arguably more fundamental than $N_\mathrm{GC}$, but requires additional assumptions regarding GC mass-to-light ratios. The \gcnumberhalomass{} relation is also approximately linear, with small scatter.} over several orders of magnitude in $M_\mathrm{h}$, with $\lesssim 1 $~dex scatter (\citealt{Harris_2014, Harris_2015}; see also \citealt[][]{Hudson2014}).
The concept of a linear $N_\mathrm{GC}$--$M_\mathrm{h}$ relation has helped to clarify aspects of the earlier literature, which focused on the more complex relationships between $N_\mathrm{GC}$ and $M_\star$ for different subsets of the galaxy population \citep[e.g.][and refs.\ therein]{Georgiev:2010aa}. These relationships with stellar mass (or luminosity) can now be interpreted as consequences of the relatively well-understood (but, again, nontrivial) $M_\star-M_\mathrm{h}$ relation \citep[e.g.][]{Guo:2010aa,Behroozi2010ApJ_SHMR_uncertainties,Moster2013}. 

Several recent studies have sought to extend observations of the
$N_\mathrm{GC}$--$M_\mathrm{h}$ to very low mass systems, comparable to the faintest dwarf galaxies in the Local Group \citep{Forbes2018,Zaritsky_2022,Carlsten22}. Constraining the relation in this regime (in which extrapolation from more massive systems predicts less than one GC per galaxy) is particularly important, because the hierarchical accumulation of clusters in larger systems is similar for a wide variety of different cluster formation scenarios \citep{ElBadry:2019, Burkert:2020aa, Nate2020, Zaritsky_2022}. In addition to the conventional scenario for cluster formation introduced above, a \textit{primordial} population of clusters may have formed prior to cosmic reionization (at approximately the same time as the first galaxies) when largely neutral gas accumulated rapidly in the most overdense dark matter halos. This scenario was proposed (in an early form) by \citet{PeeblesDicke1968ApJ} and \citet{Peebles1984}. It has subsequently been developed alongside the understanding of structure formation in a $\Lambda$CDM cosmogony \citep[e.g.][]{Forbes:1997aa,Cote:1998aa,Kruijssen_Cooper_2012_initial_mass_GCs}. Dwarf galaxies may be the best probes of this primordial scenario, because the merging and hierarchical aspects of the conventional picture are (arguably) less important in dwarf galaxies, and the conditions necessary for conventional GC formation are harder to obtain at low virial mass\footnote{More exotic scenarios for GC have been proposed, e.g.\ \citet{NaozNarayan2014ApJ}, in which GCs form outside DM halos.}.

The $M_\mathrm{GC}$--$M_\mathrm{h}$ (or $N_\mathrm{GC}$--$M_\mathrm{h}$ ) relation is therefore an extremely important constraint on models of GC formation: arguably as important as the $M_\mathrm{\star}$--$M_\mathrm{h}$ relation is for models of galaxy formation \citep[e.g.][]{Guo:2010aa}. A variety of theoretical models have been developed, which track the formation and destruction of GCs in a modern cosmological context, based on semi-analytic \citep[e.g.][]{Peng:2008aa, Boylan-Kolchin2017, Creasey_2018, ElBadry:2019, Choksi:2019aa,Chen:2023aa,De-Lucia:2024aa} and hydrodynamical methods \citep[e.g][]{EMOSAICS_2018MNRAS,Nate2020, ReinaCampos2022arXiv220411861R}. All these models have been broadly successful in reproducing linear $M_\mathrm{GC}$--$M_\mathrm{h}$ relations, particularly at the scale of the Milky Way \citep[for recent summaries see e.g.][and further discussion below]{Choksi:2019aa,Nate2020,Chen:2023aa}. However, they all involve prescriptions that require some degree of calibration, for which the \citetalias{Harris_2017} dataset has become arguably the most important point of reference.

The most significant limitation of the \citetalias{Harris_2017} dataset is that it was assembled from a large number of individual studies, each using a different method of identifying GCs and based on imaging with different depth and coverage. Almost all those studies corrected their observed counts for magnitude and area incompleteness, but not necessarily in the same way. Moreover, all the contributing studies selected samples of host galaxies from specific populations; in particular, as we show below, massive early-type galaxies are over-represented in the sample, as are members of massive clusters. Much of the earlier observational data may also be biased towards galaxies known to have significant globular cluster populations, and may under-represent systems with few or no GCs\footnote{This bias may be particularly acute for datasets based on Hubble Space Telescope observations, which provide the most accurate GC counts for individual systems.}. 

Although the factors shaping the $M_\mathrm{GC}$--$M_\mathrm{h}$ relation remain poorly understood, the relation itself provides a practical means of estimating $M_{h}$ from $N_{GC}$, a straightforward observable \citep{Spitler:2009aa}. If sufficiently accurate, such estimates of $M_{h}$ could constrain other aspects of the galaxy-halo connection. It is increasingly common for the empirical $M_\mathrm{GC}$--$M_\mathrm{h}$ relation to be used in this way \citep[e.g.][]{Beasley:2016ab,Burkert:2020aa,Forbes:2024aa}. Given  this, and the significance of the relation for theoretical models, it is worthwhile to explore new approaches to constraining $M_\mathrm{GC}$--$M_\mathrm{h}$.

In this paper we present a new measurement of the \gcnumberhalomass{} relation, which we construct by combining an all-sky imaging dataset, the DESI Legacy Imaging Survey \citep[DESI-LS;][]{Dey19_DESILegacysurvey}, with the well-established (but by no means standardized) `background subtraction' method for counting GCs around individual galaxies (as used in early work, for example, by \citealt[][]{Miller_1998, Lotz_2004, Miller_Lotz_2007} and more recently by e.g. \citealt{Lim_2018, Hughes2021_GC_NGC5128, Carlsten22, Pan2022_GC_M81,Buzzo2022_detectGC_SPLUS,Buzzo2023_GC_LSS_NGC1052}).  
Individual cluster counts determined in this way have low signal-to-noise. We therefore focus on average GC counts in bins of host galaxy luminosity (which we can associate, on average, with characteristic values of host galaxy stellar mass and virial mass). We construct these average counts using a stacking method, similar to that used by \citet{Carlsten22} and in studies of the average luminosity function and radial distributions of dwarf satellite galaxies \citep[e.g][]{Guo12_exceednumber,Wang:2012aa}. 

The main advantage of DESI-LS for this work is its significantly greater depth compared to other surveys of a similar area (in particular, the Sloan Digital Sky Survey). This makes it possible to carry out a more homogeneous statistical survey of GC systems around nearby galaxies to a depth comparable to earlier galaxy-by-galaxy studies. Here we concentrate on galaxies within $30$ Mpc drawn from the Siena Galaxy Atlas \citep[SGA,][]{Moustakas2023SGA}, a value-added photometric re-reduction of galaxies with large angular size in DESI-LS. Although the analysis we describe in this paper is less detailed than many of the individual studies compiled by \citetalias{Harris_2017}, we argue that it demonstrates the value of current and near-future all-sky surveys for such work, and offers a promising route towards further improvement in sample size and statistical homogeneity.

This paper is organized as follows. In Section \ref{sec:primary}, we introduce a sample of isolated primary galaxies  drawn from the SGA catalog. We discuss our photometric selection criteria for GC candidates in Section \ref{sec: GC selection}. Our procedure for constructing stacked GC projected density profiles is presented in Section \ref{sec: density profile}, and the profiles themselves are presented in Section~\ref{sec: observations}. In Section \ref{sec:results} we compare the profiles and corresponding GC abundances to other observations in the literature. In Section~\ref{sec: discussion}, we interpret these results, discuss the limitations of our method, and consider possible future work. We summarize and conclude in Section \ref{sec: conclusion}. Throughout, we assume a standard flat $\mathrm{\Lambda CDM}$ cosmogony with Hubble parameter $H_{0} = 67.3\,\mathrm{km\,s^{-1}\,Mpc^{-1}}$ \citep{Planck2014_cosmological_parameters}. 
    
\section{Primary galaxy selection} 
\label{sec:primary}
    The following sections describe our primary sample of isolated central galaxies drawn from the Siena Galaxy Atlas (SGA) 2020 catalog.

    \subsection{The Siena Galaxy Atlas}\label{subsec_SGA_catalog}
    The Siena Galaxy Atlas \citep{Moustakas2023SGA} is a multi-wavelength catalog of photometric properties for 383,620 nearby galaxies, based on data from the DESI Legacy Imaging Survey \citep[][DESI-LS]{Dey19_DESILegacysurvey}. DESI-LS observed $\sim14,000$ square degrees of the sky in three optical bands, $g$, $r$, and $z$, in order to identify targets for the Dark Energy Spectroscopic Instrument (DESI) survey. The survey comprises two components, referred to as the ``North" and ``South" catalogs. The South component was observed using the Dark Energy Camera (DECam) at CTIO, as the DECam Legacy Survey (DECaLS). The North component was observed by the Beijing-Arizona Sky Survey (BASS) and the Mayall z-band Legacy Survey (MzLS). Consequently, there are differences in detail between the photometric systems of the two components. The North and South catalogs have a broad region of overlap.
    
    The SGA selects galaxies with large angular size from DESI-LS for more detailed surface brightness profile analysis than was carried out by the standard DESI-LS pipeline. SGA provides 
    measurements of size, morphology and surface brightness
    in the $g$, $r$, and $z$ bands.
    SGA also defines galaxy groups, based on angular separation. A galaxy group in SGA comprises all galaxies with overlapping elliptical isophotes of diameter $2 D_{25}$ (with $D_{25}$ the length of the major axis of an ellipse corresponding to the $25 \ \mathrm{ mag\,arcsec^{-2}}$ surface brightness contour in the B-band, as reported by HyperLEDA). 
    The central galaxy of a group is taken to be the member with the largest angular size.
   
    Our primary sample is selected based on luminosity. Moreover, our GC selection includes a criterion based on cluster absolute magnitude, and we construct cluster radial density profiles in physical units. We therefore require accurate proper distances for the galaxies in our primary sample. 
    For this purpose, we match SGA to the most recent version of the Extragalactic Distance Database \citep[EDD;][]{Tully:2009aa} based on data from the \textit{Cosmicflows-4} project \citep{Tully2023ApJ_COSMICFLOWS4-galaxydistance}. EDD is a compilation of redshift-independent distances obtained with a variety of methods\footnote{Fundamental plane, Tully-Fisher, type Ia supernovae, type II supernovae, surface brightness fluctuations, the Cepheid period–luminosity relation and the tip of the red giant branch.}.  $30,502$ SGA galaxies have EDD distances ($\sim8\%$ of the SGA catalog).

    \subsection{Primary galaxy selection criteria}
    \label{sec:primary_criteria}

    We use the following criteria to select our primary SGA-EDD galaxy sample. Field names refer to the SGA catalog.

     \begin{enumerate}
        \item \texttt{GROUP\_PRIMARY = True}, to select only central galaxies in galaxy groups (as defined by SGA).

       \item \texttt{ELLIPSEBIT = 0}; this is a bitmask; a value of zero indicates the galaxy was fit well by the SGA ellipse model.

        \item  \texttt{Z\_LEDA > 0}. This criterion excludes galaxies that do not have a reliable redshift in SGA.

        \item  \texttt{(G,R,Z)\_COG\_PARAMS\_MTOT  != -1};  this criterion excludes galaxies for which SGA does not report a total apparent magnitude (in any of the three bands) based on the  curve of growth from an ellipse fit to the surface brightness profile. Very few galaxies in the  SGA catalog fail this criterion, and all those that do are excluded from our sample by other criteria. However, we also exclude NGC 5476, because the $r$-band magnitude reported by SGA seems to be spurious.

        \item Distance $0 < d_\mathrm{EDD} < 30\,\mathrm{Mpc}$, where $d_\mathrm{EDD}$ is the mean distance reported by EDD (i.e.\ an average, where multiple redshift-independent distance estimates are available). In our fiducial analysis we only include SGA galaxies that have EDD distances\footnote{In appendix~\ref{sec:appendix_sample} we report results based on Hubble-flow distances for SGA primaries that meet our other selection criteria.}. The limiting distance (30~Mpc, distance modulus $\mu \simeq 32.4$ mag) is chosen such that we expect to detect $50\%$ of the GCs around the most distant galaxy in our sample (assuming detections are limited only by magnitude, and also assuming a Gaussian GCLF; see below). We take this fiducial limiting depth to be $g = 25.5$ mag. For comparison, the typical $5\sigma$ point source depth for a source with 3 $g$-band exposures from DESI-LS (specifically, DECaLS) is $g_\mathrm{lim,5\sigma} \simeq 25$ mag\footnote{\small \url{https://www.legacysurvey.org/dr9/description/##depths}}. 
        
        We  use this global estimate of the limiting depth only to determine the $30\,\mathrm{Mpc}$ distance limit for the primary sample. To estimate GC counts for each primary, we make use of sources fainter than $g_\mathrm{lim,5\sigma}$ and different estimates of depth for different regions of each galaxy (see below).

        \item Luminosity in the $z$-band $8 \leq \log_{10}L/\lsun \leq 11.5$. This luminosity is computed from the SGA total $z$-band apparent magnitude at the distance of the primary galaxy (from EDD, as above).

     \end{enumerate}

At this point, prior to applying the additional isolation criteria described in the next section, galaxies that meet all of the above criteria are assigned to $\log_{10}$ luminosity bins of width $0.5$~dex, as shown in Fig.~\ref{fig: primary selection}.

\subsubsection{Isolation criteria} 
\label{sec:isolation_critera}

The GC systems of bright galaxies in physical (as opposed to visual) pairs or groups may overlap on the sky, complicating the measurement of stacked density profiles. On the other hand, visual neighbors with distances much larger than our limit of 30~Mpc, and dwarf satellite galaxies, are less problematic, because they contribute very few (or no) true clusters. 

The `search radius' parameter of our stacking method, $r_{search}$ (defined below) corresponds to the circular area (in kiloparsecs) around each primary in which we estimate counts of candidate GCs. This radius is scaled to the physical size of the galaxy: in each primary luminosity bin, we compute the mean physical $D_{26}$ isophotal diameters of the galaxies in kiloparsecs (converting their angular diameters into physical sizes, 
assuming their EDD distances); the corresponding radius is $r_{26} = \frac{1}{2}D_{26}$. We then define a single physical search radius for all primary galaxies in a luminosity bin, $r_\mathrm{search} = 4 r_{26}$ ($=2D_{26}$). This is the radius within which we search for GC candidates. 

We use a larger limiting radius, $2\,r_\mathrm{search}$, to define the outer limit of an annular region from $\,r_\mathrm{search}$ to $2\,r_\mathrm{search}$), in which we measure the background count of candidate sources.

In order to avoid issues with overlapping GC systems, we impose further criteria on our sample of primaries such that they have no neighbors likely to cause a significant excess of GC candidate counts within $2 r_\mathrm{search}$. Specifically, we exclude any primary that has one or more neighbours of similar physical luminosity and size to the primary. To do this, for each primary, we examine every apparent neighbor within $2 r_\mathrm{search}$ and apply the following criteria:

\begin{itemize}
    \item If the neighbour does not have a measured redshift, the $z$-band absolute magnitude of the neighbor must be more than 2 magnitudes fainter than that of the primary, and the diameter of the neighbor must be $D_\mathrm{26,ngb} < \frac{1}{3} D_\mathrm{26,primary}$. 

    \item If the neighbor has a measured redshift, we apply the previous magnitude and size criteria \textit{only} if the implied line-of-sight separation from the primary is $<10\, \Mpc$. If a neighbor fails either criterion (i.e. if it is as large or as bright as the primary) but it has a redshift consistent with a line-of-sight distance $>10\, \Mpc$, then we do not exclude the primary on the basis of that neighbor.

\end{itemize}

These criteria do not exclude primaries in cases where large or bright neighbors are centered just outside, but very close to, the $2\,r_\mathrm{search}$ boundary. Such cases are unlikely to contribute a significant excess of the background counts, because GC systems are typically concentrated close to their host galaxies.

\subsubsection{Restriction to the Southern DESI-LS footprint}

We use imaging data from DESI-LS data release 9 \citep[DR9;][]{Dey19_DESILegacysurvey}. Over $70\%$ of the primaries in our sample are included in the DESI-LS South catalog (see Fig.~\ref{fig: primary selection}, which shows the total number of primaries selected in each luminosity bin and the number of those in the South catalog). Given this, we only use primaries in the South catalog in this paper, to avoid the additional uncertainty involved in working with two separate photometric systems.

\subsubsection{Composition of the primary sample}

\begin{figure}

\includegraphics[width=\linewidth]{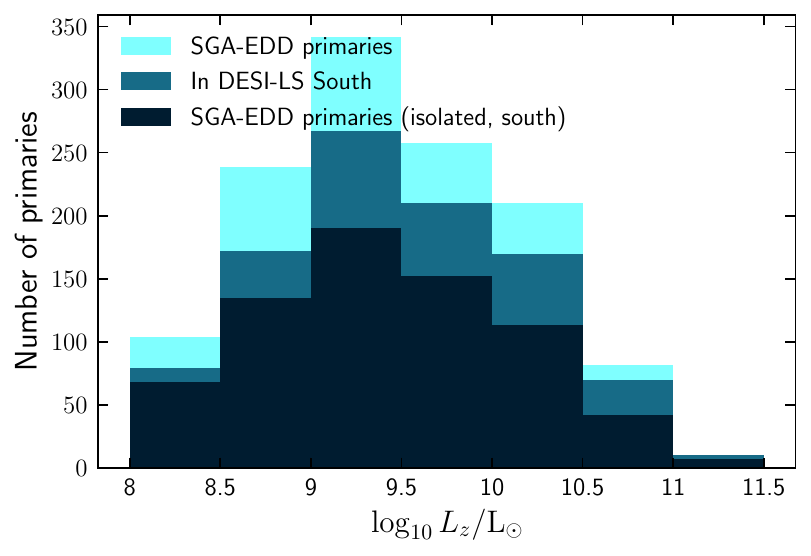}
\caption{The luminosity distribution of our primary galaxies selected from SGA (black histogram), according to the criteria in Section~\ref{sec:primary}. The cyan and teal histograms correspond to supersets of the primary sample as described in the text.
\label{fig: primary selection}}
\end{figure}

Fig.~\ref{fig: primary selection} shows the luminosity distribution of our primary sample. The cyan histogram corresponds to $1245$ SGA galaxies with EDD distances that meet all the criteria described in section~\ref{sec:primary_criteria} (i.e.\ excluding the isolation criteria). Among these, $978$ (teal histogram) are included in the DESI-LS South catalog. Our final primary sample comprises the $707$ galaxies that also meet the isolation criteria described in section~\ref{sec:isolation_critera}.

The distributions in Fig.~\ref{fig: primary selection} show a deficit of low and high mass galaxies relative to the field luminosity function. The lack of faint primaries is, for the most part, because we require EDD distances\footnote{We examine this bias in  appendix~\ref{sec:appendix_sample}, using Hubble-flow distances for a larger sample of galaxies from SGA.}. The majority of EDD distance measurements for isolated low-mass galaxies are based on the Tully-Fisher method, which requires a sufficiently coherent and massive disk, increasingly uncommon at lower masses.

\begin{figure*}
    \includegraphics[width=\linewidth]{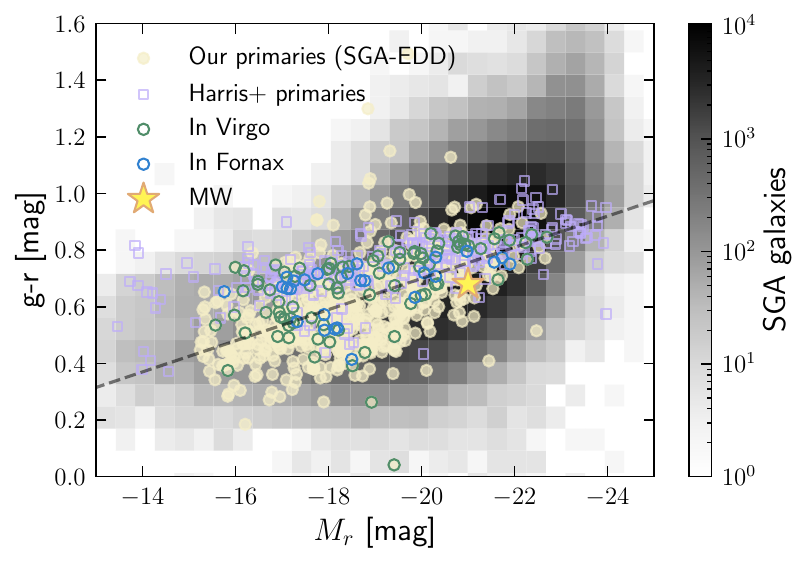}
    \caption{The color -- magnitude diagram of our primary sample (yellow circles), plotted over the distribution of all SGA galaxies (grey histogram). Colors are corrected for MW reddening, but not for reddening intrinsic to each galaxy, hence the population of extremely red galaxies at high luminosity; dusty star-forming galaxies therefore contribute to the population of very red galaxies with high luminosity. The dashed line shows a fiducial separation between blue cloud and red sequence galaxies, $(g-r) = -0.055 M_r-0.4$, as in \citet{Liao2023_colorgradient_galaxies}.  Primaries in our sample that are members of the Virgo and Fornax clusters are marked with green and blue open circles, respectively. We also show the distribution of galaxies from the catalog of \citet[][purple squares; many galaxies in this catalog are not in our sample of SGA primaries]{Harris2013}. The star symbol shows the estimated color and luminosity of the Milky Way from \citet{Licquia:2015wn}.
    \label{fig: color-mag of primaries}}
\end{figure*}

Fig.~\ref{fig: color-mag of primaries} shows the color -- magnitude diagram of galaxies in our sample. The dashed line shows a fiducial separation of red sequence and blue cloud galaxies in the DESI-LS bands, $(g - r) = -0.055 M_r - 0.4$ \citep[as used by][]{Liao2023_colorgradient_galaxies}. Relative to the full SGA catalog, our criteria favor the selection of primaries in the `green valley' and the redder part of the blue cloud. They exclude both brighter and bluer galaxies, and the cluster red sequence. This is most likely the result of our isolation criterion, which removes most cluster galaxies (although see appendix~\ref{sec:appendix_sample}) and bright $L_{\star}$ galaxies in loose groups. However, as a green-valley system, the Milky Way is within the population of massive galaxies included in our sample \citep{Licquia:2015wn}. This is helpful, because the MW provides the most complete and least ambiguous data for a single GC system.

There are few extremely luminous galaxies in our sample volume. These are typically brightest cluster galaxies (BCGs). Our sample volume includes two massive clusters, Virgo and Fornax. Our criteria select 85 and 27 primary galaxies within $2\,\Mpc$ of the approximate centers of Virgo and Fornax, respectively. 

We also show the distribution of galaxies from the \citetalias{Harris_2017} catalog in Fig.~\ref{fig: color-mag of primaries}. The Harris et al.\ galaxies mostly follow the red sequence at lower masses, reflecting the large number of GC counts originating from deep surveys of Virgo and Fornax members. From this figure we conclude that neither our sample nor that of \citetalias{Harris_2017} is fully representative of the overall galaxy population, although our sample is less biased at lower luminosity. We return to this point when we discuss our results in section~\ref{sec: discussion}. We discuss the bias of our sample with respect to the full SGA catalog further in appendix~\ref{sec:appendix_sample}.

\section{Globular cluster selection} 
\label{sec: GC selection}
 
After defining a sample of primaries, we identify all sources around those primaries that are potential GCs. This section describes our criteria to distinguish GC candidates from foreground stars, background galaxies and spurious sources.

\subsection{Search and background regions}

\begin{figure}
    \includegraphics[width=\linewidth]{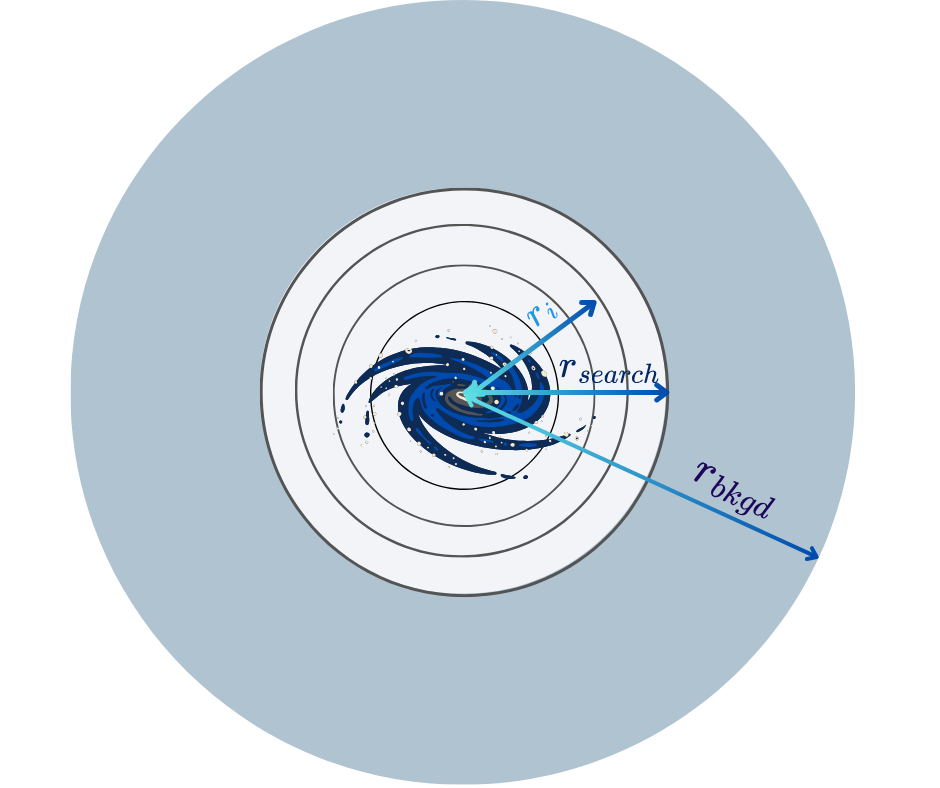}
    \caption{Illustration of our definition of the search radius, $r_\mathrm{search} = 2 D_{26} \ [\kpc{}]$, and background radius,  $r_\mathrm{bkgd} = 2 r_\mathrm{search}$,  around a primary. We also illustrate the division of the area within the search radius into annuli (of radius $r_{i}$) for the purpose of constructing a surface density profile. The number of annuli and the scale of $r_\mathrm{search}$ relative to the central galaxy are chosen for illustration, and hence are not to scale. \label{fig:search_radius}}
\end{figure}

As illustrated in Fig.~\ref{fig:search_radius}, we define a circular region around each primary of radius $r_{search} = 4 \, r_{26} (=  2 \, D_{26})$ 
(these quantities are defined in section~\ref{sec:isolation_critera}). This is the region within which we aim to identify and count candidate GCs. To statistically correct those counts for contamination, we define a larger annulus of radius $r_\mathrm{bkgd} = 2\, r_\mathrm{search} (=8r_{26})$. Objects identified as candidates in the annulus $r_\mathrm{search} < r < r_{bkgd}$  are used to establish a local background level.

\subsection{GC candidate selection criteria}

Having identified all DESI-LS DR9 sources within $r_\mathrm{bkgd}$ around each primary, we apply the following criteria to pick out likely GC candidates.

\begin{figure} 
    \includegraphics[width=\linewidth]{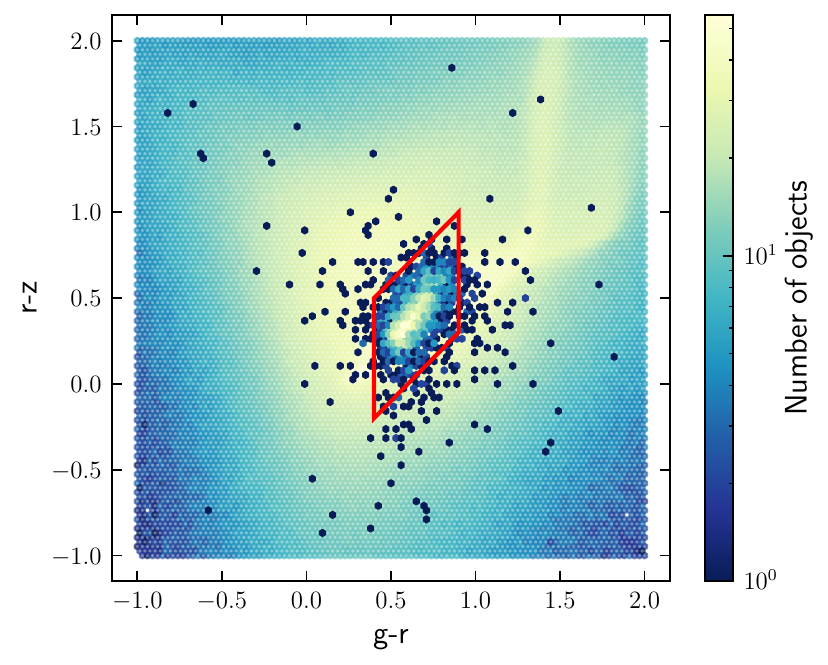}
    \caption{Color-color diagram of sources around our selected primaries in the DESI-LS bands, over-plotting (with the same color scheme) the density of confirmed GCs around SLUGGS hosts 
    \citep{Brodie_2014_SLUGGS_survey}. 
    The color bar shows the counts of objects in each color-color bin. The red rhomboid region, defined in the text, is our GC candidate color selection, which encloses $89\%$ of the SLUGGS GCs.
    \label{fig: GC candidates color selection}}
\end{figure}

\begin{enumerate}
    
    \item We include all objects that are not matched to an entry in the Gaia DR2 \citep{2018GAIA_DR2} catalog (\texttt{REF\_ID = 0}). For all objects that are matched to Gaia, we include only those with \textit{both} low proper motion and small parallax $(\sqrt{\texttt{PMRA}^2 + \texttt{PMDEC}^2}< 5 \ [\mathrm{mas/yr}]\, \texttt{\&}\, \texttt{PARALLAX} < 3\sigma + 0.3 \ [\mathrm{mas}])$. These criteria exclude nearby stars.
    
    \item $\texttt{PSFSIZE\_G} < 2\arcsec$ and $\texttt{SHAPE\_R} < 2\arcsec$. This selects objects that are not significantly extended according to either of these DESI-LS measures of apparent size.
    
    \item Following \citet{Hahn_DESI_BGS}, we separate GC candidates (and other point sources) from galaxies by requiring $G_\mathrm{Gaia} - r_{raw} < 0.6$, where $G_{\mathrm{Gaia}}$ is the magnitude in the Gaia G-band and $r_{raw}$ is the `raw' (without correction for Milky Way transmission) magnitude in the DESI-LS r-band. Gaia measures magnitudes of sources assuming a diffraction-limited PSF, whereas the DESI-LS magnitude captures a greater fraction of light from extended sources. Sources with only a small difference between these two magnitudes are likely to be point sources.

    \item \texttt{FLUX\_G},  \texttt{FLUX\_R} and  \texttt{FLUX\_Z} $>0$; \texttt{NOBS\_G}, \texttt{NOBS\_R} and \texttt{NOBS\_Z} $> 0$. These are quality cuts, requiring that the source is detected in all three DESI-LS bands and has at least one observation in each band.
    
    \item In the ($g-r$, $r-z$) color space (Fig.~\ref{fig: GC candidates color selection}) we require candidates to be in the trapezoidal region defined by $0.4< g-r <0.9$ and the upper and lower boundaries  $r-z < (g-r) + 0.1$ and $r-z > (g-r) - 0.6$. This region is motivated by the colors of confirmed GCs reported in the literature, specifically the SLUGGS survey 
    \citep{Brodie_2014_SLUGGS_survey}.
    SLUGGS observed in the  Subaru/Suprime-cam $g$, $r$ and $i$ bands; we obtain DESI-LS colors for these objects by matching the coordinates of GCs in SLUGGS to DESI-LS sources with a tolerance of $2 \arcsec$.
    As shown in the figure, this region contains 89\% 
    of the SLUGGS clusters. The region overlaps the Milky Way stellar locus around the main sequence turnoff, but is slightly offset from it in $r-z$.

    \item  Following \citet{Carlsten22}, we assume that the GCLF follows a Gaussian distribution with a mean $\mu = -7.2$ in the $g$-band and a standard deviation of $\sigma = 0.7$. We only select sources having absolute magnitudes within $\pm 3\sigma$ of the mean of this assumed LF. We compute absolute magnitude of a source assuming the same distance as its corresponding primary\footnote{For simplicity we use the same GCLF parameters for all galaxies. \citet{Jordan_2007_GCLF} show evidence that, in the Virgo cluster, the GCLF is broader in more luminous hosts. We find that assuming a much broader GCLF ($\sigma=1.3$) or a brighter mean ($\mu=-8$) for brighter primaries has only a minor effect on the stacked profiles that we show in Section \ref{sec: observations}.}.
    
\end{enumerate}

\section{Globular cluster density profile} 
\label{sec: density profile}

Having identified a set of GC candidates around a given primary, we compute a background subtracted profile of excess candidate counts in circular annuli, $\Sigma(r_i)$, following (for example) \citet{Guo12_exceednumber}:

\begin{align}
\begin{split}
 \Sigma(r_i) &= \frac{N(r_i)}{A(r_i)} - \frac{N_{bkgd}}{A_{bkgd}}\\
 &= \frac{N(r_i)}{\pi (r^2_i - r^2_{i-1})} - \frac{N_{bkgd}}{\pi (r_{bkgd}^2-r^2_{search})},
\end{split}
\end{align}
where $N(r_i)$ is the number of GC candidates in the $i^\mathrm{th}$ annulus and $A(r_i)$ is the area of that annulus. We define the background count, $N_\mathrm{bkgd}$, to be the number  of GC candidates in the annulus $r_\mathrm{search} < r < r_\mathrm{bkgd}$, with $A_\mathrm{bkgd} = \pi (r^2_\mathrm{bkgd} - r^2_\mathrm{search})$ the area of that annulus. We use 10 radial bins to compute the profile in the region $r < r_\mathrm{search}$. The search radius, profile annuli and background radius are illustrated in Fig.~\ref{fig:search_radius}.

\subsection{Surface-brightness dependent background subtraction}  

The increase in the surface brightness of the primary galaxy at smaller galactocentric radii makes it relatively less likely that fainter sources will be detected there. To first approximation, this will reduce the count of both true clusters and background sources by equal amounts. Subtracting a constant background level from all annuli would result in an over-subtraction for inner annuli, which have brighter detection limits.  We therefore determine a separate background count for each annulus: we find the apparent magnitude of the faintest source visible in an annulus, $m_\mathrm{min,i}$, and then use the number of background objects brighter than $m_\mathrm{min,i}$ as the background count for that annulus\footnote{The DESI-LS pipeline reports the $5\sigma$ point source detection magnitude at the location of each source. We also considered the average of this quantity over a given annulus as an estimate of the liming depth. However, $m_\mathrm{min}$ gives an estimate that corresponds more closely to the shape of the central galaxy surface brightness profile, and tends to the background count in the outer annuli. The average PSF depth, in contrast, is approximately constant except in the innermost few annuli; it is also somewhat brighter than $m_\mathrm{min}$ (by 1-2 mag) even in the background region, and hence gives a rather conservative estimate of the detection limit.}.

This correction assumes that the detection efficiency of candidates drops from 100\% to zero at $m_\mathrm{min}$, both in the search region and in the background. This is unlikely to be true in detail, especially for annuli with bright   $m_\mathrm{min}$. If the detection efficiency falls more gradually with source magnitude in such regions than it does in the background, the correction above would be an underestimate.

\subsection{GCLF correction for faint clusters}  \label{sec: corrections}

We further correct the background-subtracted count in each annulus 
to account for the portion of the GCLF fainter than the limiting apparent magnitude $m_\mathrm{min,i}$. As above, we assume a universal Gaussian GCLF. The cumulative value of GC candidates at a given threshold absolute magnitude $M_\mathrm{min,i}$ is therefore given by
\begin{align}\label{eq: cdf GCLF}
    \mathrm{CDF}(M_\mathrm{min,i}) = 0.5  \left[1 + \mathrm{erf}\left(\frac{M_\mathrm{min,i} - \mu}{\sqrt{2} \sigma}\right)\right],
\end{align}
with the error function $\mathrm{erf} (z) = \frac{2}{\sqrt{\pi}}\int_0^z e^{-t^2}dt$. From the CDF, we compute the fraction of GCs brighter than $M_\mathrm{min,i}$, then estimate the total GC count in the annulus by multiplying the background-subtracted density by a factor
\begin{align}
    f_\mathrm{GCLF} = \frac{1}{\mathrm{CDF}(M_\mathrm{min,i})}.
\end{align}

\section{Stacked GC density profiles} 
\label{sec: observations}

Following the above procedure, we estimate the GC candidate number density profiles around each galaxy in our fiducial SGA-EDD sample.
We divide the primaries into $7$ luminosity bins of width $0.5$~dex over the range $\log_{10} L_{z}/\lsun = [8, 11.5]$. The number of primaries in each bin ranges from 7 (in the highest-luminosity bin) to 190 (in the bin $\log_{10} L_{z}/\lsun = [9, 9.5]$). 

\begin{figure} 
\includegraphics[width=\linewidth]{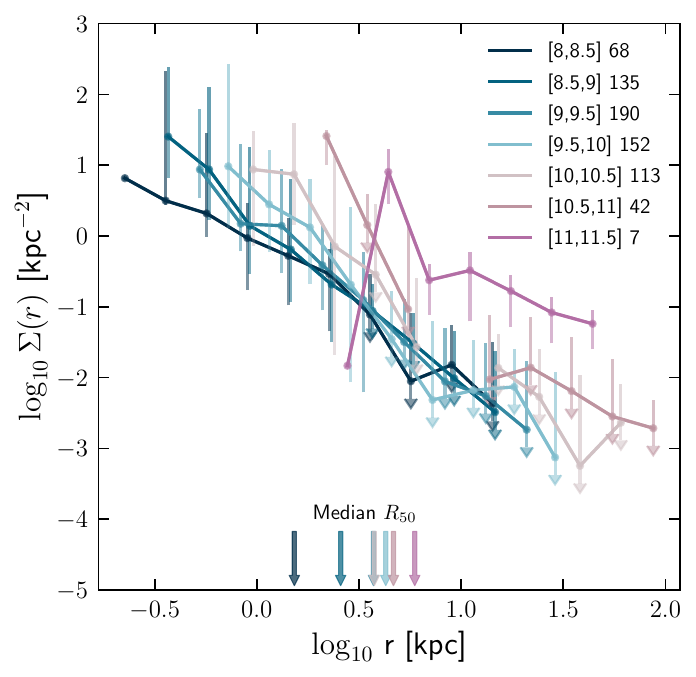}
\caption{The median of GC candidate surface density profiles stacked in logarithmic bins of primary luminosity, for our fiducial SGA-EDD sample. The error bars show the 16 -- 84 percentile range in each bin; downward-pointing arrows indicate that the minimum estimated count is consistent with zero. The range of log luminosity for each bin and the corresponding number of galaxies are shown in the legend. The arrows mark the median half-light radius of primaries in each luminosity bin.}
\label{fig: GC profiles}
\end{figure}

Fig.~\ref{fig: GC profiles} shows the `stacked median' profile for galaxies in each luminosity bin, averaging the individual counts for each galaxy in each annulus.
The profiles of less luminous primaries, over the range $8 
< \log_{10} L_{z}/\lsun < 10$, have almost the same shape: the GC candidate density is high ($\Sigma = 1 - 10\,\kpc^{-2}$) near the center and rapidly decreases at larger radii, falling to between $\sim10^{-3}$ and $10^{-4}\,\kpc^{-2}$ at $r_\mathrm{search}$. A trend of increasing GC candidate density with primary luminosity is apparent for luminosities $\log_{10} L_{z}/\lsun > 9.5$, particularly outside $r\sim 10$~kpc. The inner cut-off of each profile and the gaps apparent in the profiles for the higher luminosity bins are discussed below.

The individual profiles that contribute to these stacks are noisy. Their large variance may be due in part to the intrinsic, physical scatter between the GC profiles of different galaxies, but in this case we suspect it is dominated by radially-dependent detection effects. These affect many individual galaxies in the stack in a similar way (because, on average, galaxies with similar luminosity have similar surface brightness profiles) and can therefore be imprinted on the average profiles. 
In the following subsection, we briefly describe the origin of these effects and the practical steps that we take to mitigate them when constructing the stacks.

\subsection{Systematic effects}
\label{sec:profile_systematics}

In individual profiles, some annuli may have spuriously low (or less often, high) signal due to local variations in the accuracy of the DESI-LS photometry (see also section~\ref{sec:completeness_corrections_discussion}). Typically, these effects are more severe at smaller galactocentric radius, where the surface brightness of the host is higher. For very luminous galaxies, these effects are clearly apparent when overplotting DESI-LS sources on the images, as shown in appendix~\ref{sec:massive_galaxy_effects}.

The stacked average would be biased if we included annuli with severe photometric artifacts. However, low or negative counts alone are not a reliable way to identify those annuli. Counts fluctuating around zero are `normal' for the background-dominated outer annuli. Ignoring all annuli with negative counts would therefore introduce a significant bias in the outer region of the profiles, and would also fail to eliminate spuriously small positive counts in the inner regions. 

Instead, to mitigate this, we only include a given annulus from a given primary in the stacked average if the number of GC candidates in that annulus satisfies the condition $|N(r_{i}) - \hat{N}(r_{i})| < 5\sigma$, where $\hat{N}(r_{i})$ is the mean count for the annulus taken across all galaxies in the stack, and $\sigma = \sqrt{\hat{N}(r_{i})}$ is the Poisson error. 

The rationale for this is to identify annuli that deviate significantly from the stacked average, using the Poisson fluctuation of the background as a measure of significance. This is far from an ideal way to identify outliers (not least because it relies on the `raw' stacked average) but we find it eliminates the worst cases while still allowing outer annuli with statistically `reasonable' negative counts to contribute, such that the excess count at large radius goes smoothly to zero.

Excluding the most significant `outlier' annuli in this way does not prevent all spurious systematic effects from being imprinted onto the average profile. In particular, in the brightest luminosity stacks ($\log_{10} L_z/\lsun = [10, 10.5], [10.5, 11], [11, 11.5]$), we detect almost no GC candidates in the innermost annuli ($r < 1$--$3 \ \kpc$) around most individual galaxies. This is most obvious in the highest luminosity stack, $\log_{10} L_z/\lsun = [11, 11.5]$, where there is no signal in the stacked profile up to $r \sim 3 \ \kpc$. This is a consequence of the very high central surface brightness in these primaries, which renders even the brightest GCs undetectable at the spatial resolution of the DESI-LS images. 

Any candidates that are detected in this inner region (which may themselves be less reliable, see section~\ref{sec: discussion}) are up-weighted strongly by the GCLF correction described in Sec.~\ref{sec: corrections}. This may overestimate the total density. Overall, however, the stacked average is dominated by primaries with near-zero detections, leading to a inner cut-off in the profiles that occurs at larger radii in more luminous galaxies.

Essentially the same effect creates `gaps' around $r\sim10$~kpc in the two bins that span the range $10< \log_{10} L / \mathrm{L_{{\odot}}}<11$. Although the surface brightness of the primaries at $r<10$~kpc is not high enough to prevent detection of the brightest globular clusters, the use of single-component models of the primary galaxy in the DESI-LS photometric pipeline can lead to regions of over-subtraction at larger radius, and artifacts associated with galactic structure \citep[see section 3.2.2 of][see also our appendix A]{Moustakas2023SGA}.

The gaps and cut-offs in the stacks are a fair reflection of the typical behavior seen for individual galaxies. As a rough rule of thumb, these effects become noticeable in individual profiles around the galaxy half-light radius, $R_{50}$. The average $R_{50}$ in each luminosity bin is shown by the arrows in Fig.~\ref{fig: GC profiles}. In principle, a decline in the density of GCs in the centers of galaxies may not be entirely spurious -- there may be physical reasons to expect such a decline in the central few kiloparsecs of very massive galaxies. However, we suspect that detection effects dominate the counts in the inner regions of our stacked profiles, not least through comparison to more thorough observations of individual examples in the literature (see appendix~\ref{sec:massive_galaxy_effects}). 

Zero or greatly underestimated counts due to a very bright (or poorly determined) detection limit should not be treated in the same way as `simple' non-detection of the faint tail of the GCLF in regions with much lower surface brightness. Low but positive counts at small radii (or in regions with severe photometric issues) should not be corrected for using the method in section~\ref{sec: corrections}, because those objects are likely to be artifacts, and the true surface brightness limit is not known. When stacking, those low or negative counts should not contribute to the stacked average in the same way as `genuine' fluctuations around zero signal in the outer regions of galaxies, where GC candidate luminosities are relatively robust and hence  the GCLF correction is justified. However, we have not found a justifiable way to determine when a low or negative count in a bin is `genuine' and when it is `spurious'. As noted above, our $5\sigma$ `filtering' of the stack only partly mitigates this problem. A more statistically robust treatment of these issues is clearly important for future applications of this method. Here we simply accept that our GC candidate counts may be over- or underestimates (likely the latter) in the center of galaxies, more so in the higher luminosity bins; and that some stacked profiles will have gaps for the reason given above. In spite of these problems, we believe the resulting profiles are sufficiently representative to proceed with a first analysis of the DESI-LS data. Our analysis in the following sections will focus on the lower luminosity bins. Our results for the highest luminous bins, which could have much greater uncertainty, are nevertheless useful for comparison with the many previous studies of GC counts for more massive galaxies in the literature.

The exception is the highest luminosity bin $\log_{10} L_z/\lsun = [11, 11.5]$ which contains only 7 galaxies, at least 4 of which have severe photometric issues as described above. The uncertainty associate with this small sample is compounded by a potential `Eddington bias' due to the difficulty of measuring accurate stellar masses for very massive galaxies with large angular size: a handful of fainter galaxies with overestimated luminosities may make a spurious contribution to the sample in this bin. Likewise a significant number of galaxies that should be in this bin may have their luminosities underestimated. For this reason, we do not expect that the results in this bin are at all robust, although we show them in subsequent figures. We discuss individual galaxies in this highest luminosity bin further in Appendix~\ref{sec:massive_galaxy_effects}.

\subsection{Total cluster counts}

We obtain an average total number of GCs for each bin of primary luminosity by integration under the stacked average profile. As described in the previous section, in the three highest-luminosity bins, these integrated counts are strongly affected by systematics in the source photometry. To be conservative, we do not interpolate across gaps in the profiles of the most massive luminosity bins (i.e.\ gaps  contribute zero GCs)\footnote{We find interpolating linearly across the gaps makes only a few per cent difference to the counts.}. Counts in regions close to gaps are also likely underestimated due to the effects described above. We do not make any correction for this underestimate.

\begin{figure}
    \includegraphics[width=\linewidth]{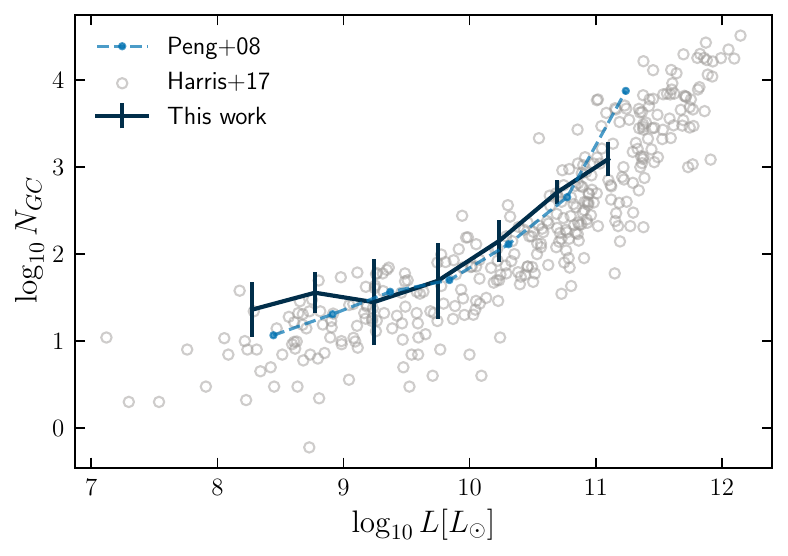}
    \caption{The relationship between average luminosity and GC number obtained from our stacked profiles, compared to the average relation obtained for Virgo cluster members by \citet{Peng2008_GC_earlytypeVirgo} and individual points from the compilation of \citetalias{Harris_2017}. Note that luminosities for our results and those of \citet{Peng2008_GC_earlytypeVirgo} are in the $z$ band and those for the \citetalias{Harris_2017} are in the $K$ band; we do not correct for the difference in this figure (see text).
    \label{fig: Luminosity - N_GC}}
\end{figure}

Fig.~\ref{fig: Luminosity - N_GC} shows the average counts we obtain in each luminosity bin, compared to data from \citet[][for galaxies in the Virgo cluster]{Peng2008_GC_earlytypeVirgo} and the compilation of \citetalias{Harris_2017}. Our results show a very similar trend of counts with luminosity to both of these datasets, and hence also agree with recent models that reproduce the same trend \citep{Choksi:2019aa,Chen:2023aa}. Our results also show similar amplitude for this relation, with marginal evidence slightly for higher amplitude in the lowest two luminosity bins. In the highest luminosity bin, our count is lower than the other two datasets; this is expected given the small number of galaxies in this bin and the systematic effects described above (highly incomplete profiles due to high surface brightness and sensitivity to under- or over-estimated luminosity for individual galaxies due to the small sample size).

The \citet{Peng2008_GC_earlytypeVirgo} and \citetalias{Harris_2017} datasets have slightly different amplitudes in this figure. There are several possible reasons for this. The \citet{Peng2008_GC_earlytypeVirgo} data comprise mostly early types in a dense environment. At fixed stellar mass, early types are likely to have higher virial mass, which may lead to a higher cluster count; we discuss this further below. Moreover, these datasets refer to different passbands ($z$ in the case of \citeauthor{Peng2008_GC_earlytypeVirgo} and our work, and $K$ in the case of \citeauthor{Harris_2017}). An approximate correction for this difference is $\log_{10}\,L_K \approx \log_{10}\,L_z + 0.3$ for a galaxy with $g-r=0.8$ \citep[][]{Bell:2003aa}, which would bring the mean of the three datasets into rough agreement. We return to this point below in our discussion of the relation between $N_\mathrm{GC}$ and stellar mass.

\section{Results}
\label{sec:results}

\begin{figure}
    \includegraphics[width=\linewidth]{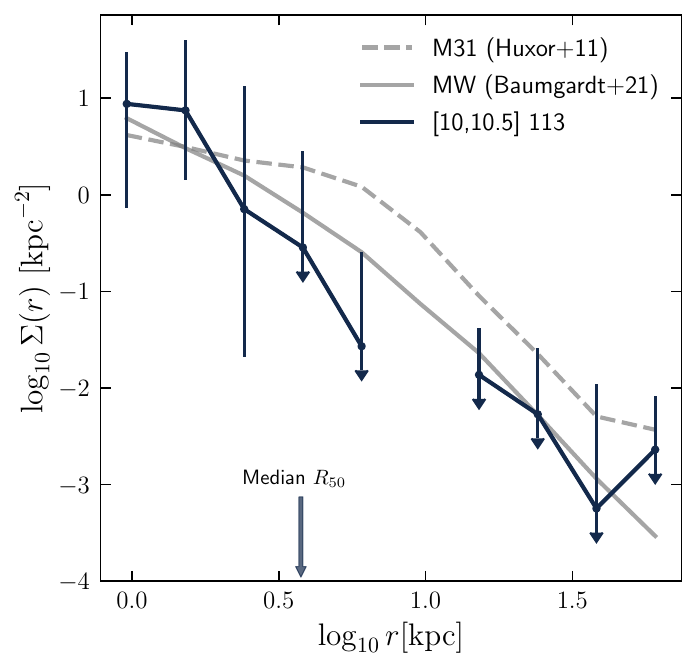}
    \caption{Median GC candidate density profile of 
    primaries with luminosity $10 < \log_{10}L/\lsun < 10.5$ (black line)
    compared to the GC density profiles of the MW \citep[solid grey line,][]{Baumgardt2021_GC_MW_profile} and M31 \citep[dashed grey line,][]{Huxor11_GC_M31}.  
    \label{fig: E and S GC profiles, bin 5}}
\end{figure}

\subsection{Overall profile amplitude and shape}

Fig.~\ref{fig: E and S GC profiles, bin 5} compares our stacked profile for $\log_{10} L \in [10, 10.5]$ to the density profile of Milky Way GCs, based on data from \citet{Baumgardt2021_GC_MW_profile}, and M31 GCs, from \citet{Huxor11_GC_M31}. The Milky Way profile is an average of the three profiles obtained by projecting the three-dimensional cluster positions along each axis of the Galactocentric XYZ coordinate system.

The shape and amplitude of our profile are in good agreement with that of the Milky Way, which is arguably the most complete and robust GC dataset available. The underestimation of counts in our data from $3-10$~kpc relative to the MW profile is evident. The Milky Way is likely at the upper end of the luminosity range to which we compare \citep{Licquia:2015wn}. M31 may be significantly more luminous, and appears to have substantially more recent accretion than the Milky Way, perhaps explaining the higher amplitude\footnote{
Most of the additional GCs in the more recent M31 dataset are in the outskirts, $\log_{10} r \ge 1.5 \ \kpc{}$, and hence do not change the profile reported by \citet{Huxor11_GC_M31} significantly.} of its GC profile \citep[e.g.][]{Dey:2023aa}.

\begin{figure}
    \includegraphics[width=\linewidth]{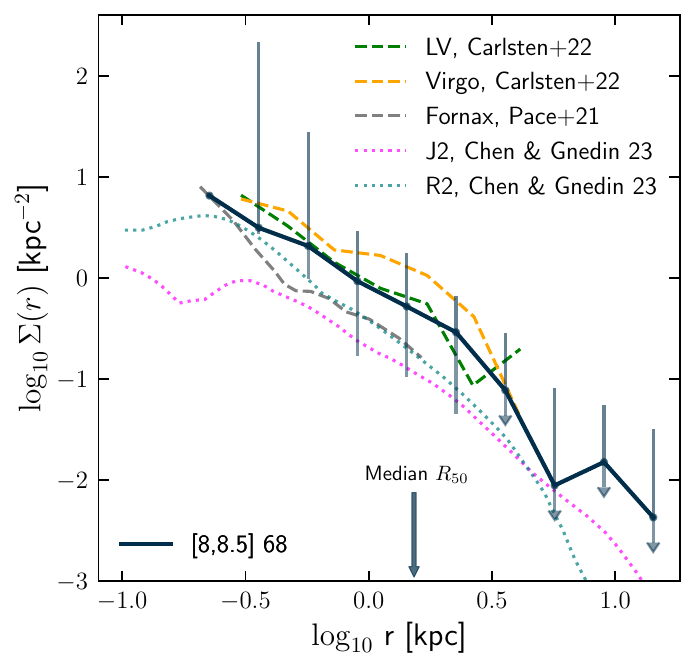}
    \caption{Median GC candidate density profile of 
    primaries with $8 < \log_{10}L/\lsun < 8.5$ (black line)
    compared to the GC density profiles of the Fornax dwarf spheroidal \citep[grey dashed line,][]{Pace_2021}; simulations of galaxies analogous to the Fornax dSph \citep[magenta and teal dotted lines,][]{Chen:2023aa}; and observations of Local Volume and Virgo dwarf galaxies \citep[green and orange dashed lines, respectively][]{Carlsten22}. 
    \label{fig: GC profiles, bin 1}}
\end{figure}

Fig.~\ref{fig: GC profiles, bin 1} shows the median GC profile for primaries in the luminosity range $8 < \log_{10} L/\lsun < 8.5$ (solid black line). In this luminosity range, the \citet{Into_Portinari10_color_mass_to_light_relations} CMLR gives $M/L \approx 1$, such that the corresponding mass range for our sample is approximately
$8< \log_{10} M_{\star}/\msun < 8.5$. Our result is consistent with the stacked average found by \citet{Carlsten22} for dwarf galaxies in Local Volume with masses $5.5 < \log_{10} M_{\star}/\msun < 8.5$ (green dashed line), although our least massive primaries correspond to the upper mass limit of their sample. The stacked average found by \citet{Carlsten22} for dwarf galaxies in Virgo cluster (orange dashed line) is higher than our result, but within our estimated uncertainty. \citet{Carlsten22} argue the difference in average counts between their Local Volume and Virgo samples is evidence for a systematically higher abundance of dwarf galaxies at fixed stellar mass in dense environments, as suggested in earlier work \citep[e.g][]{Peng:2008aa,Lim_2018}. Our result is also consistent with, although somewhat higher than, observations of GCs around the Fornax dwarf spheroidal galaxy \citep[$M_\star\approx2 \times 10^7\, \msun{}$; data from][]{Pace_2021} 
and simulated analogues of that galaxy \citep[][]{Chen:2023aa}. The higher amplitude is perhaps not surprising, as the average stellar mass of our sample in this range is roughly an order of magnitude greater than than mass of the Fornax dwarf.

Cosmological models predict that GC populations comprise two components with different origins and hence different density profiles: a compact component corresponding to clusters formed `in situ', and an extended component, corresponding to clusters accreted from tidally disrupted satellites \citep[][]{Belokurov:2023aa, Chen:2024aa}. We do not see clear signs of inflections indicative of multiple components in our profiles, perhaps in part because of the limitations of our data at small radii in bright galaxies; we do see marginal evidence of a more rapid increase in amplitude at radii $\gtrsim10\,\mathrm{kpc}$ with increasing luminosity.

\subsection{Number of globular clusters as a function of host halo mass}

\begin{figure*}
\centering
\includegraphics[width=0.9\linewidth]{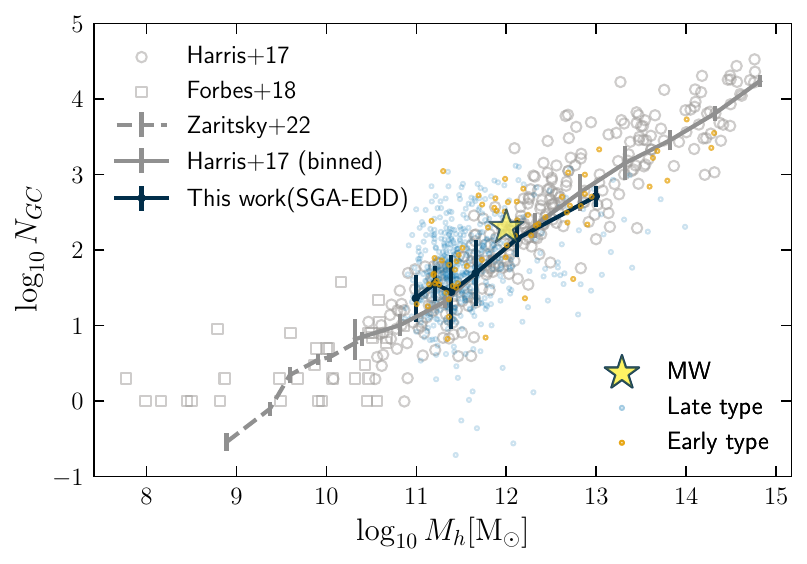}
\caption{Our result of the \gcnumberhalomass{} relation (black line), obtained from the median GC candidate count in each luminosity bin and plotted at the average halo mass for galaxies in the corresponding bin (see text). The error bars indicate the median absolute deviation. Each orange/blue circle represents the number of GCs of an individual primary in this work, colored based on the type of the primary. The open circles and solid grey line show observational data and the corresponding mean relation from \citetalias{Harris_2017}. At low masses, the squares and dashed grey line show data for dwarf galaxies from \citet{Forbes2018} and \citet{Zaritsky_2022}, respectively. The GC count and halo mass of the Milky Way are shown by the yellow star.} 
\label{fig: NGC-Mh}
\end{figure*}

To compare with the \citetalias{Harris_2017} \gcnumberhalomass{} relation, we estimate a host DM halo mass for each primary using the stellar mass -- halo mass (SMHM) relation of \citet{Moster2018MNRAS}. To obtain the stellar mass of each primary, we use the `exponential' color-mass-to-light relation (CMLR) given by \citet{Into_Portinari10_color_mass_to_light_relations},
\begin{equation}
    \log_{10} M_\star/L_z = 1.158 (g - r) - 0.619.
    \label{eq:ip}
\end{equation}
This CMLR relation was calibrated for primaries with color\footnote{For this, we use the \texttt{COG\_PARAMS\_MTOT} SGA magnitudes.} $0.1<(g-r)<0.85$. For galaxies colors above or below these limits, we assume $M_{\star}/L$ is constant and equal to the value of Eq.~\ref{eq:ip} at either $(g-r)=0.85$ or  $(g-r)=0.1$, respectively.
We then convert the average stellar mass of primaries in a given bin to a representative halo mass using the mean relation of \citet{Moster2018MNRAS}. 
The inferred range of halo mass for our sample is approximately $10.5 \lesssim \log_{10} M_h/\msun \lesssim 13.5$. The full uncertainties in $M_{\star}/L$ are not easy to estimate, but are probably $\gtrsim0.2$~dex, with estimates becoming less reliable for nearby galaxies with large angular size (i.e.\ in the presence of well-resolved color gradients and internal structure).
The difficulty of measuring stellar masses to an accuracy $\lesssim 0.1$~dex also contributes to uncertainty in the average SMHM relation \citep[e.g.][]{Behroozi2010ApJ_SHMR_uncertainties}. We consider the effects of these uncertainties below.

Fig.~\ref{fig: NGC-Mh} shows the \gcnumberhalomass{} relation we obtain from the stacked average in each luminosity bin (black line). We compare this to data from \citetalias{Harris_2017} and  \citet{Zaritsky_2022}, for the least massive dwarf galaxies. At the high-mass end of our sample, our results are consistent with \citetalias{Harris_2017}. At the low-mass end, the average GC candidate counts we infer are higher than those of \citetalias{Harris_2017}. There are several possible reasons for this: a systematic floor in our counts due to our background subtraction method (i.e.\ `contamination' of our counts); a difference in the nature of the objects considered as GC candidates; differences in the method used to correct counts for magnitude and area incompleteness; or a genuine difference, perhaps due to different underlying samples of host galaxies. We examine each of these possibilities in the following section.

We also show in Fig.~\ref{fig: NGC-Mh} the scatter of individual $N_{GC}$ measurements for each individual galaxy in our sample. These are inevitably very noisy compared to the stacked average. Nevertheless, their dispersion is broadly consistent with the scatter of the \citetalias{Harris_2017} data. The scatter in these galaxy-by-galaxy counts is likely to be somewhat larger than the uncertainty we associate with the average count derived from the stacks. When computing stacked average, we exclude the contribution of a radial bin from an individual galaxy if it is more than $5\sigma$ from the mean count in that bin over all galaxies in the stack (see section~\ref{sec:profile_systematics}). We do this in order to reduce the effect of outliers on the stacked average. However, we do not exclude these (potentially) extreme values when computing the individual total counts galaxy by galaxy\footnote{The $5\sigma$ criterion has other limitations. In the innermost annuli ($r < R_{50}$), only  $3-5 \%$ of galaxies have any detected counts. The average is correspondingly low and $\sigma$ large, hence it is impossible to distinguish outliers reliably (or at all) by this criterion. At larger radii, most galaxies have candidates detected. However, in the outermost annuli, the scatter can be large, such that only $15-35\%$ of galaxies have counts within $5\sigma$.}.

At the lowest virial masses, we see a tail towards higher $N_{GC}$ that is not apparent in \citetalias{Harris_2017}, although it is consistent with the scatter of the \citetalias{Harris_2017} data at higher virial mass. This tail contributes to the higher average $N_{GC}$ we find in the lowest virial mass bins.

Uncertainty in primary stellar masses of galaxies, and uncertainty in the mapping those stellar masses to virial mass, may give rise to a difference between the `average' $N_{GC}$ values obtained by integrating the stacked average profile over a wide bin of stellar mass, and the average that would be obtained from individual measurements of $N_{GC}$ in a narrow range of stellar mass.

We use a simple Monte-Carlo model to investigate this qualitatively, as follows. We assume a fiducial \gcnumberhalomass{} relation following the linear fit reported in \citetalias{Harris_2017}, with Gaussian scatter in cluster number at a fixed halo mass parameterized by a dispersion, $\sigma_N= 0.26 $~dex, 
independent of $M_{h}$. We draw random halo masses for galaxies from a log-normal distribution matched very approximately to the distribution of halo mass inferred for our sample (mean $10^{11.5}\,\mathrm{\msun}$, dispersion $0.5$~dex). For each halo, we determine $N_{GC}$ by sampling from our fiducial relation\footnote{The low-mass tail of this Gaussian extends below the lowest halo mass in our SGA-EDD sample.}. We obtain corresponding stellar masses, $M_\star$, by sampling from the \citet{Moster2018MNRAS} Gaussian SMHM relation\footnote{For simplicity, we assume no additional uncertainty due to the conversion between luminosity and stellar mass.}. We then bin the model galaxies by their stellar mass, compute the \gcnumberhalomass{} relation from these averaged data, and compare to the underlying SMHM relation of our toy model. We generate 100 realizations, each comprising 700 halos. 

\begin{figure}
    \includegraphics[width=\linewidth]{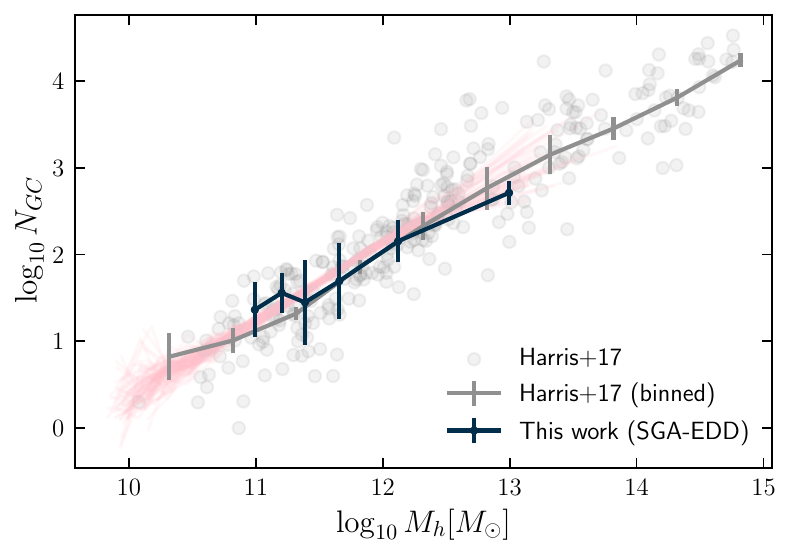}
    \caption{The pink lines show the distribution of 100 Monte-Carlo samples from a simple model of the \gcnumberhalomass{} relation, based on the linear fit to \gcnumberhalomass{} given in \citetalias{Harris_2017} and a Gaussian distribution of halo mass.  We compare these realizations to the data shown in Fig.~\ref{fig: NGC-Mh}. Our stacked averages are consistent with the Monte Carlo distribution even in the lowest mass bins where they appear marginally inconsistent with the average of the \citetalias{Harris_2017} data in the same range of $M_\mathrm{h}$ (see text). 
    \label{fig: mc sampling of Ngc-Mh}}
\end{figure}

Fig.~\ref{fig: mc sampling of Ngc-Mh} shows the resulting average $N_{GC}$, which is is biased towards slightly higher $N_{GC}$ than the input relation, and has substantial scatter at the limits of the $M_\mathrm{h}$ range.

These differences result from the scatter in $M_\star$ at fixed $M_\mathrm{h}$. In the halo mass range where we see an apparent excess in $N_{GC}$ relative to the (per mass bin) average of the \citetalias{Harris_2017} data, that average is itself below the expectation of the linear fit to the full dataset reported in \citetalias{Harris_2017}. Moreover, although our data are still somewhat higher than the average predicted by the \citetalias{Harris_2017} linear fit, they are within the plausible range of variation for the Monte-Carlo samples we draw from it, particularly given the greater variation due the smaller samples at the low and high extremes of $M_\mathrm{h}$. We conclude that our results are in good agreement with those of \citetalias{Harris_2017}. Including other effects, such as the uncertainty in the \citetalias{Harris_2017} linear fit, would increase the scatter of our Monte Carlo realizations even further, and hence reduce the significance of the disagreement between our results and those of \citetalias{Harris_2017} even further.

\section{Discussion: GC counts in low-mass dark matter halos} 
\label{sec: discussion}

We have shown that it is possible to obtain meaningful constraints on GC counts across a wide range of halo mass using only the DESI-LS pipeline photometry. We now briefly assess how the limitations of this photometry and our stacking approach might influence our results. We concentrate on our finding that the average number of GCs in systems with halo mass $10.5 < \log_{10} M_h/\msun < 12$ (i.e.\ in halos less massive than that of the Milky Way) is somewhat higher than that of galaxies in the same mass range reported in \citetalias{Harris_2017}, as shown in Fig.~\ref{fig: NGC-Mh}. Although the statistical significance of this difference in the averages may be low (as we argue above), we also see larger scatter than the \citetalias{Harris_2017} data in this regime. Therefore, taking this result at face value, in the following subsections we discuss possible reasons for a difference in GC candidate counts at low masses relative to \citetalias{Harris_2017}, first considering sources of contamination, then the completeness, and, finally, possible differences in the underlying galaxy samples.

\subsection{Spurious sources} 

It is likely that our counts are contaminated to some degree by entirely spurious sources, or sources with  inaccurate photometry, resulting from limitations of the Legacy Survey reduction in regions close to very bright or extended galaxies  \citep[see][for further discussion of the DESI-LS  pipeline in this regime]{Moustakas2023SGA}.

Our selection criteria include quality cuts to remove pipeline artifacts, the most obvious of which only appear in a single band or have extreme colors. Spurious sources that survive these cuts may include deblending artifacts (i.e.\ light from parts of the primary wrongly modeled as separate sources) and foreground stars (or background galaxies) that have poor photometry because their light is blended with light from the primary (or other sources in the field). In principle, this blending effect is distinct from the background subtraction issues that lead to higher incompleteness of genuine clusters and background sources in the centers of bright galaxies. However, it may be closely related, if more severe over- or under- subtraction of the galaxy light profile affects the subsequent source detection and modeling stages of the pipeline. These types of spurious source would naturally be concentrated around primary galaxies, and  their density may change with galactocentric radius (to first order, declining with the surface brightness of the primary, but perhaps influenced by features such as spiral arms)\footnote{This effect is illustrated for massive galaxies by the example in appendix~\ref{sec:massive_galaxy_effects}.}. 

Therefore, the higher counts we find in low mass systems may represent a systematic floor at which spurious sources selected by our GC candidate criteria dominate over genuine GCs. We have inspected sources identified as candidates around a randomly chosen subset of low-mass primaries; in appendix~\ref{sec:low_mass_galaxy_effects} we show some examples from the lowest luminosity bin. In contrast to the effects seen in more massive galaxies, we see no evidence of problems with the DESI-LS model of the primary or other obvious artifacts in these low-mass systems. 
However, in several cases, we see $\lesssim10$ candidates in the visual body of the galaxy; these may be clusters of some form, but their photometry may suffer from crowding and contamination by the diffuse light of the galaxy. Based on this visual inspection, a floor of $\gtrsim20-30$ spurious sources per galaxy on average seems unlikely, though not impossible. A more likely possibility is that even one or two spurious (or even genuine) sources may be associated with large surface brightness correction factors, if the central surface brightness of the dwarf is sufficiently high.
Although not all galaxies have sources detected in high surface brightness regions, those that do may inflate the scatter in $N_{GC}$. The underlying DESI-LS photometry in bright regions of galaxies could likely be improved, and more detailed analysis undertaken to limit the contribution of poorly-photometered sources to GC counts. Such work would be valuable for subsequent applications of this method.

\subsection{Genuine non-GC sources associated with primaries}
\label{subsection:real_but_not_gc_sources}

There may be populations of genuine sources associated with the primary galaxies that meet our candidate selection criteria, and hence contribute to an excess of source counts, but which do not meet the conventional definition of GCs. Given our restriction to unresolved sources in a narrow color range (which should exclude HII regions, for example), such sources are perhaps most likely to be young or otherwise unusual massive star clusters \citep{Holtzman_1992, Whitmore_1999, PortegiesZwart2010_YMC, Longmore:2014aa, Bastian_2016_YMC} 
Where such sources are unresolved (as in the DESI-LS images), it is highly unlikely that they could be distinguished reliably from the spurious sources discussed in the previous section. A further complication is that, as recent work has suggested, the number of low-luminosity GCs in some galaxies may be larger than expected under the standard assumption of a universal Gaussian GCLF \citep[e.g.][]{Huxor:2014aa, Carlsten22, Floyd_2024}.

It is currently ambiguous whether such clusters should be counted in surveys seeking to constrain the \gcnumberhalomass{} relationship, even if they can be correctly identified. It is similarly unclear whether they have been counted in any of the datasets used by \citetalias{Harris_2017}. This issue is perhaps more acute for actively star-forming dwarf galaxies, in which bright young clusters 
are relatively common.

\subsection{Completeness corrections}
\label{sec:completeness_corrections_discussion}

GC counts are typically corrected for magnitude incompleteness (i.e.\ detection sensitivity) assuming a universal Gaussian GCLF, as described in section~\ref{sec: corrections}. It may not be completely beyond doubt that the GCLF is universal, or Gaussian \citep[e.g.][]{Carlsten22}; these assumptions are also somewhat dependent on the operational definition of GCs. Nevertheless, such corrections are common in the literature. The uncertainty in our results due to this assumption is likely comparable to that associated with the \citetalias{Harris_2017} datasets in the same respect. 

GC counts are also usually (though not always) corrected for incomplete coverage of the virial radius of the host galaxy. This correction is likely to vary more significantly among the \citetalias{Harris_2017} datasets. For example, of the few datasets in \citetalias{Harris_2017} that includes low-mass late-types in the field is that of \citet{DiNino_2009}; however, this study focused exclusively on the pseudo-bulges of its target galaxies, using the Hubble Space Telescope. The corresponding coverage is therefore much smaller than used in our work. To our knowledge, the counts reported in \citetalias{Harris_2017} are not corrected for that limited coverage.

At face value, area completeness is not a factor in our results, because our data extend uniformly to a well-defined galactocentric distance that is, by design, much larger than the likely extent of the GC population in each primary luminosity bin. As discussed above, there are effectively unobserved regions in the inner parts of many of our sample galaxies, where the high background surface brightness and DESI-LS reduction artifacts results in non-detections for the majority of galaxies in a given stack. This incompleteness is greatest in the brighter primary luminosity bins (see section~\ref{sec:profile_systematics} and appendix~\ref{sec:massive_galaxy_effects}).

In the lowest luminosity bins, we have uniform and presumably substantially complete (albeit low purity) coverage to our limiting magnitude. It is therefore possible that the higher counts for low mass galaxies seen in Fig.~\ref{fig: NGC-Mh} are partly the result of probing larger galactocentric radii than usual for such systems. In the lowest luminosity bins, for which the median $R_{50}\approx 1.5 - 2.5\,\mathrm{kpc}$, between 10 and 30\% of GC candidate counts in our sample come from the region beyond $2R_{50}$. 
This is approximately consistent with the estimate made by \citet{Carlsten22} that $\sim1/3$ of the total GC candidate count for a typical object in their dwarf galaxy sample is associated with clusters beyond $2R_{50}$.
These numbers are necessarily rather uncertain given very low typical counts of $N_{GC}\lesssim10$ and the potential for increasing scatter in the SMHM relation at low masses. At present, we can only speculate that field galaxies in this mass range may not have been studied in sufficient number or to sufficient distance to identify their more distant clusters routinely.

\subsection{Differences in the primary sample}
\label{sec:discussion_differences_in_primary_sample}

The final possibility we consider is that the excess is due to differences between the population of galaxies  in our low mass sample and those in the \citetalias{Harris_2017} compilation in the same range of halo mass. Such differences are possible: we have shown, in Fig.~~\ref{fig: color-mag of primaries}, that sub-MW mass galaxies in \citetalias{Harris_2017} are mostly on the red sequence, and that our sample includes a much larger fraction of blue cloud galaxies at lower mass. 

\citet{Carlsten22} discuss evidence for environmental effects on GC populations in detail, in particular the possibilitiy that GCs may be more abundant in low-mass cluster members than field dwarfs at fixed stellar mass (they find twice as many galaxies with $N_{GC}>10$ in Virgo compared to the Local Volume). A significant proportion of the \citetalias{Harris_2017} sample at low stellar mass are cluster members \citep[e.g.\ those from][]{Peng:2008aa}, whereas our sample is biased towards isolated galaxies. Therefore, at face value, the slightly higher counts we find at low halo mass relative to \citetalias{Harris_2017} are the opposite of what would be expected.

However, the difference suggested by \citet{Carlsten22}, though significant, is small in absolute terms, and likely well within the uncertainties of our counts, those of \citetalias{Harris_2017} and those associated with the mapping of stellar mass to halo mass. Moreover, the \citet{Carlsten22} sample is also not fully representative of the low-mass field (it includes a substantial number of satellites, for example), and has a somewhat lower average stellar mass than our sample.

In particular, redder field galaxies of low-to-intermediate mass are not well represented in any previous study. Our isolation criterion means concentrates most of our sample around the green valley at the Milky Way stellar mass, and the requirement of distances in the EDD catalog tend to restrict the sample to the green valley also at lower masses (yellow points in figure \ref{fig: color-mag of primaries}; we discuss the effect of the EDD requirement below). In contrast, the low-mass galaxies in the \citetalias{Harris_2017} sample lie mostly on the cluster red sequence (purple squares in Fig.~\ref{fig: color-mag of primaries}). To examine the effect of this difference, we constructed a subset of low-mass galaxies in our sample with a similar color-magnitude range to the \citetalias{Harris_2017} data in the same mass range. We find no significant difference between the median GC profile of the Harris-like subset and that of our full sample; the excess at low mass is also seen in our result for the Harris-like subset. We conclude that the cluster red sequence and  passive field galaxies at low-to-intermediate galaxy masses have, on average, similar GC abundances.

The requirement of EDD distances for our sample galaxies favors redder galaxies at lower masses (in particular, cluster members with fundamental plane or surface brightness fluctuation distances). As described in appendix~\ref{sec:appendix_sample}, we have also examined a sample obtained without this requirement, using Hubble-flow distances instead. We call this sample SGA-HF. It is much larger than our SGA-EDD sample, particularly in the lowest luminosity bins: a nine-fold increase in the sample size in the  $8.0 < \log_{10} L/\lsun < 8.5$ bin and a four-fold increase in the $8.5 < \log_{10} L/\lsun < 9$ bin (see Fig.~\ref{fig: primary sample no EDD} in appendix~\ref{sec:appendix_sample}). This provides more representative coverage of the color-magnitude distribution of galaxies less luminous than the MW in the full SGA catalog. The GC abundances we obtain from this SGA-HF sample are systematically lower than those for our SGA-EDD sample, albeit consistent within $1\sigma$ (see Fig.~\ref{fig: NGC-Mh relation SGA sample} in appendix~\ref{sec:appendix_sample}). They are therefore closer to the direct average of \citetalias{Harris_2017}, although the interpretation of this result is not straightforward; we discuss this further in appendix~\ref{sec:appendix_sample}.

At the MW stellar mass and above, the isolation criterion `pinches' the distribution of galaxies in our sample more strongly towards the green valley, and excludes a significant population of galaxies in loose groups and MW-mass galaxies with massive satellites, which tend to have higher star formation rates. Further work is therefore required to determine whether the GC populations of such galaxies are different from those of massive galaxies in our sample. However, the MW itself is not representative of this population; it likely falls in the green valley \citep{Licquia:2015wn} and hence should be represented well by our sample.

We have already remarked on the possibility of Eddington bias affecting the highest luminosity bins. Since the requirement of EDD distances greatly limits the number of galaxies in our sample at low masses compared to galaxies of intermediate luminosity (as shown in Fig.~\ref{fig: primary selection}; compare Fig.~\ref{fig: primary sample no EDD} in appendix~\ref{sec:appendix_sample}), it is possible that a `reverse' Eddington bias operates here also, i.e.\ that galaxies with under-estimated luminosity and sizable GC populations contaminate lower luminosity bins with intrinsically small numbers of primaries in SGA-EDD. However, this effect seems unlikely to dominate the counts at the level observed, and we see no obvious sign of substantial contamination by brighter galaxies in visual inspection of examples from the lowest luminosity bins (see appendix~\ref{sec:low_mass_galaxy_effects}).

Finally, we note that population-level effects like these, although complex, are naturally accounted for by recent models of cluster formation in a cosmological context, such as \citet{EMOSAICS_2018MNRAS}, \citet{Chen:2024aa} and \citet{De-Lucia:2024aa}. These models predict counts that are broadly in line with the \citetalias{Harris_2017} average. Further investigation of the origin of the scatter predicted by these models, and predictions for subsets of the galaxy population with exceptionally high or low counts, as in \citet{De-Lucia:2024aa}, would be extremely useful.

\section{Conclusion} 
\label{sec: conclusion}

The sky coverage and sensitivity of the DESI Legacy Imaging Survey \citep{Dey19_DESILegacysurvey} is sufficient to extend statistical methods for estimating globular cluster counts from ground-based photometry  (specifically, the subtraction of a nearby background of candidate GC sources to derive excess counts close to galaxies) to a large fraction of the nearby galaxy population. Here, we have presented a first attempt to construct a large, homogeneous sample of GC systems around isolated galaxies of luminosity $10^8 < L < 10^{11.5}\, \lsun$ at distances $<30 \ \Mpc$. From this sample, we provide a new estimate of the relationship between GC number and virial mass (or stellar mass), which can be compared to other compilations of literature data such as that of \citet[][]{Harris_2017}. This approach can address the question of whether the more heterogenous data underlying previous measurements of this relation are sufficiently representative. 
It also allows us to investigate variations among subsets of galaxies, such as early and late types.

Our sample in this paper comprises 707 galaxies, drawn from the Siena Galaxy Atlas \citep{Moustakas2023SGA}. We stack these galaxies in bins of luminosity to obtain average GC counts as a function of host luminosity and virial mass. We find the following results:

    \begin{enumerate}
        \item The GC density profile around Milky Way-like primaries in our sample corresponds approximately to the GC density profile of the MW itself.

        \item The amplitudes of our average GC density profiles are similar across luminosity bins at small galactocentric radii, but increase systematically with host luminosity in the outer regions of galaxies (beyond 10~kpc).

        \item The \gcnumberhalomass{} relation  we obtain is broadly consistent with that of \citetalias{Harris_2017}. Our result therefore provides an independent validation of the relation obtained by \citetalias{Harris_2017} across a wide range of halo mass for isolated galaxies and further evidence that this relation, overall, is not strongly affected by sample selection effects.

        \item  We find a marginal excess of total GC number at halo masses $M_\mathrm{h} < 10^{12}\,\msun$, relative to  \citetalias{Harris_2017}. A significant tail of individual galaxies with high $N_{GC}$ in this mass range, not present in the \citetalias{Harris_2017} dataset, contribute to this excess. The excess is consistent with the expected variance of a sample drawn from a linear \gcnumberhalomass{} fit to the \citetalias{Harris_2017} data (see also the concluding paragraph, below).

    \end{enumerate} 

Our findings broadly support the reliability of our GC selection criteria and background subtraction approach (which are similar to those used in many previous photometric studies of smaller samples) when applied to large photometric surveys at the depth of DESI-LS. They also provide validation of the stacking method for constructing average GC candidate profiles. Our results motivate further improvement and extension of this technique to address some limitations of the present work, including the following:

\begin{enumerate}

\item The photometry of bright galaxies in DESI-LS DR9 suffers from over-subtraction effects at intermediate galactocentric radii due to the difficulty of estimating an accurate sky background around sources with large angular size \citep[e.g.][]{Moustakas2023SGA}. This leads to radial variations of source detection efficiency and source photometry systematics. We attempt to account for the expected smooth variation in the detection efficiency for GC candidates associated with the increasing surface brightness of the primary galaxy. However, this correction may under-estimate the counts in regions of high surface brightness, because it assumes that sources brighter than the estimated magnitude limit are detected with 100\% efficiency. Moreover, our method may not work well in regions where more severe systematic effects dominate. Such cases are clearly visible in the individual and stack-average profiles of the most massive galaxies in our sample. These effects makes it difficult to compare our results directly to the extensive literature for on massive galaxies, because those galaxies have the most complex photometric systematics in DESI-LS.

\item By restricting our primary sample to a volume of radius 30~Mpc, we naturally limit the number of massive galaxies, again limiting comparison with earlier work. Since more massive galaxies have more clusters in the bright tail of the GCLF, it may be possible to use a larger distance limit for brighter primary luminosity bins and still obtain sufficient GC counts in the DESI-LS images.

\item The analysis of the stacked counts and their associated uncertainties could be improved by a more robust statistical treatment of outliers, non-detections and missing data. 

\end{enumerate}

The catalog of GC counts provided by \citetalias{Harris_2017} has become an important reference point for models of GC formation. The fact that we have obtained a very similar \gcnumberhalomass{} relation, using an independent approach with a different imaging dataset and sample selection, supports this use of the \citetalias{Harris_2017} \gcnumberhalomass{} (or \gcmasshalomass{}) relation (or fits to it).  

The only significant discrepancy we find relative to \citealt{Harris_2017} is an apparent systematic excess of average cluster counts at low halo mass.  We have considered several possible explanations, both physical and related to limitations of our data. With out present dataset, the significance of this excess is marginal. Nevertheless we believe it deserves further theoretical consideration and investigation with other datasets, because the distribution of counts in this mass range may have significant impact on the calibration of models of GC formation (for example, regarding the formation of a `primordial' cluster population in halos at high redshift). As we discuss in section~\ref{sec: discussion}, it would be interesting to investigate what changes in models of GC formation are required to increase GC counts for present-day field galaxies, within the constraints imposed by observations massive systems and the (currently very limited) data from even less massive satellites \citep{Forbes2018,Zaritsky_2022}. Models could also be used to predict variations in \gcnumberhalomass{} among different subsets of the galaxy population. On the observational side, the approach in this paper could be used to explore GC counts around low-mass galaxies further with existing data from DESI-LS and the Hyper Suprime-Cam SSP survey. In the next decade, such results will be tested definitively by observations with Euclid \citep{Euclid-Collaboration:2024aa,Euclid-Collaboration:2024ab,Kluge:2024aa,Saifollahi:2024aa}, the Roman Space Telescope \citep{Spergel:2015aa, Akeson:2019aa, Dage2023}, and the Vera C.\ Rubin telescope's LSST survey \citep{Ivezic:2019aa,Usher2023}.

A catalog of the total number of GCs and dark matter halo mass for our sample of 707 galaxies is provided at \url{https://doi.org/10.5281/zenodo.14175206}.

\section{Software} 
\label{sec:software}

Numpy \citep{Numpy_2011}, Scipy \citep{2020SciPy-NMeth}, Astropy \citep{Astropy_2013}, Matplotlib \citep{Matplotlib_2007CSE}, Pandas \citep{Pandas}, WebPlotDigitizer (\url{https://automeris.io/wpd/}). 

\section{Acknowledgments}

We thank the anonymous referee for their helpful suggestions to improve the paper. We acknowledge financial support from a Taiwan Ministry of Education (MoE) Yushan Fellowship awarded to APC, MOE-113-YSFMS-0002-001-P2, and the Taiwan National Science and Technology Council (NSTC) grants 109-2112-M-007-011-MY3, 112-2112-M-007-017-MY3 and 113-2112-M-007-009. This work used high-performance computing facilities operated by the Center for Informatics and Computation in Astronomy (CICA) at National Tsing Hua University, funded by MoE, NSTC, and National Tsing Hua University.

The Siena Galaxy Atlas was made possible by funding support from the U.S. Department of Energy, Office of Science, Office of High Energy Physics under Award Number DE-SC0020086 and from the National Science Foundation under grant AST-1616414.

The Legacy Surveys consist of three individual and complementary projects: the Dark Energy Camera Legacy Survey (DECaLS; Proposal ID \#2014B-0404; PIs: David Schlegel and Arjun Dey), the Beijing-Arizona Sky Survey (BASS; NOAO Prop. ID \#2015A-0801; PIs: Zhou Xu and Xiaohui Fan), and the Mayall z-band Legacy Survey (MzLS; Prop. ID \#2016A-0453; PI: Arjun Dey). DECaLS, BASS and MzLS together include data obtained, respectively, at the Blanco telescope, Cerro Tololo Inter-American Observatory, NSF’s NOIRLab; the Bok telescope, Steward Observatory, University of Arizona; and the Mayall telescope, Kitt Peak National Observatory, NOIRLab. Pipeline processing and analyses of the data were supported by NOIRLab and the Lawrence Berkeley National Laboratory (LBNL). The Legacy Surveys project is honored to be permitted to conduct astronomical research on Iolkam Du’ag (Kitt Peak), a mountain with particular significance to the Tohono O’odham Nation.

NOIRLab is operated by the Association of Universities for Research in Astronomy (AURA) under a cooperative agreement with the National Science Foundation. LBNL is managed by the Regents of the University of California under contract to the U.S. Department of Energy.

This project used data obtained with the Dark Energy Camera (DECam), which was constructed by the Dark Energy Survey (DES) collaboration. Funding for the DES Projects has been provided by the U.S. Department of Energy, the U.S. National Science Foundation, the Ministry of Science and Education of Spain, the Science and Technology Facilities Council of the United Kingdom, the Higher Education Funding Council for England, the National Center for Supercomputing Applications at the University of Illinois at Urbana-Champaign, the Kavli Institute of Cosmological Physics at the University of Chicago, Center for Cosmology and Astro-Particle Physics at the Ohio State University, the Mitchell Institute for Fundamental Physics and Astronomy at Texas A\&M University, Financiadora de Estudos e Projetos, Fundacao Carlos Chagas Filho de Amparo, Financiadora de Estudos e Projetos, Fundacao Carlos Chagas Filho de Amparo a Pesquisa do Estado do Rio de Janeiro, Conselho Nacional de Desenvolvimento Cientifico e Tecnologico and the Ministerio da Ciencia, Tecnologia e Inovacao, the Deutsche Forschungsgemeinschaft and the Collaborating Institutions in the Dark Energy Survey. The Collaborating Institutions are Argonne National Laboratory, the University of California at Santa Cruz, the University of Cambridge, Centro de Investigaciones Energeticas, Medioambientales y Tecnologicas-Madrid, the University of Chicago, University College London, the DES-Brazil Consortium, the University of Edinburgh, the Eidgenossische Technische Hochschule (ETH) Zurich, Fermi National Accelerator Laboratory, the University of Illinois at Urbana-Champaign, the Institut de Ciencies de l’Espai (IEEC/CSIC), the Institut de Fisica d’Altes Energies, Lawrence Berkeley National Laboratory, the Ludwig Maximilians Universitat Munchen and the associated Excellence Cluster Universe, the University of Michigan, NSF’s NOIRLab, the University of Nottingham, the Ohio State University, the University of Pennsylvania, the University of Portsmouth, SLAC National Accelerator Laboratory, Stanford University, the University of Sussex, and Texas A\&M University.

BASS is a key project of the Telescope Access Program (TAP), which has been funded by the National Astronomical Observatories of China, the Chinese Academy of Sciences (the Strategic Priority Research Program “The Emergence of Cosmological Structures” Grant \#XDB09000000), and the Special Fund for Astronomy from the Ministry of Finance. The BASS is also supported by the External Cooperation Program of Chinese Academy of Sciences (Grant \# 114A11KYSB20160057), and Chinese National Natural Science Foundation (Grant \#12120101003, \#11433005).

The Legacy Survey team makes use of data products from the Near-Earth Object Wide-field Infrared Survey Explorer (NEOWISE), which is a project of the Jet Propulsion Laboratory/California Institute of Technology. NEOWISE is funded by the National Aeronautics and Space Administration.

The Legacy Surveys imaging of the DESI footprint is supported by the Director, Office of Science, Office of High Energy Physics of the U.S. Department of Energy under Contract No. DE-AC02-05CH1123, by the National Energy Research Scientific Computing Center, a DOE Office of Science User Facility under the same contract; and by the U.S. National Science Foundation, Division of Astronomical Sciences under Contract No. AST-0950945 to NOAO.



%

\vspace{5mm}





\appendix

\section{Detection effects around massive galaxies}
\label{sec:massive_galaxy_effects}

\begin{figure}
    \includegraphics[width=\linewidth]{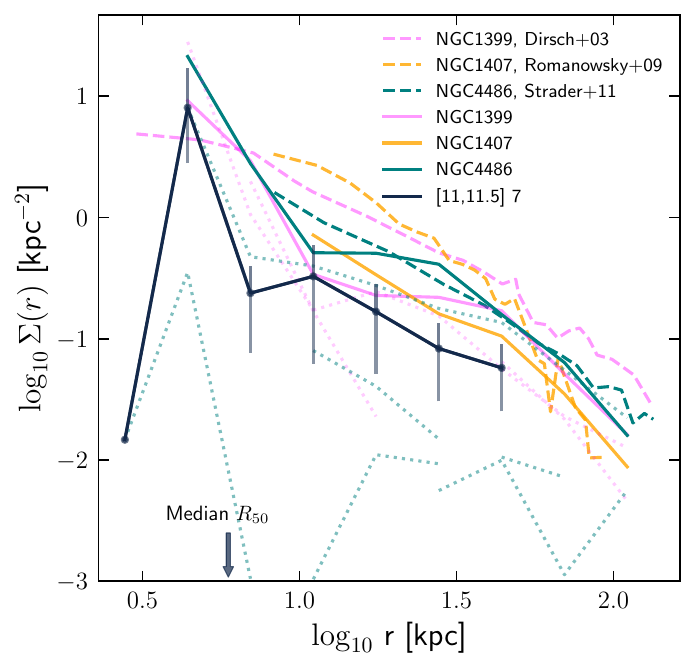}
    \caption{The median GC number density profile (black line) with MAD errorbars, for primary galaxies in the range $11 < \log_{10} L/\lsun < 11.5$. Five of the seven individual galaxies in this bin that pass our isolation criterion (and hence contribute to the average) are shown with teal dotted lines. The other two are shown with solid lines: NGC 4486/M87 (teal) and NGC 1407 (yellow). We also show, in magenta, three galaxies in this luminosity range that excluded by our isolation criteria; we highlight one of these, NGC 1399, with a solid line. 
    We compare our profiles for  NGC 4486/M87, NGC 1407 and NGC 1399 to results in the literature from \citet{Strader_2011},  \citet{Romanowsky_2009} and \citet{Dirsch2003}, respectively, using dashed lines with the same colors. 
    \label{fig: GC bin 7}}
\end{figure}

Fig.~\ref{fig: GC bin 7} shows individual GC number density profiles for primary galaxies in the luminosity range $11 < \log_{10} L/ \lsun < 11.5$, the most luminous galaxies in this study. We compare our stacked SGA-EDD median profile in this range (black line) with the individual massive galaxies: NGC 4486 (M87), NGC 1407 and  NGC 1399. NGC 1399 fails our isolation criteron and hence is not included in the SGA-EDD median profile; curiously, the other two galaxies do not fail this criterion, despite being cluster BCGs. This emphasizes the fact that the isolation criterion we impose for the purpose of establishing a `clean' background region is not directly equivalent to a typical measure of environmental density.

All three GC systems have been the subject of many previous studies. Our profile for NGC 4486 (teal solid line) is close to observations by \citet{Strader_2011} (teal dash line) based on spectroscopic confirmation of candidates identified in Subaru SuprimeCam imaging. Our profile for NGC 1407 has a lower amplitude than that reported by \citet{Romanowsky_2009}, based on Subaru SuprimeCam imaging only.
Our profile for NGC1399 also has lower amplitude than observed by \citet{Dirsch2003}, based on imaging over a $200 \ \kpc \times 200 \ \kpc$ region with the MOSAIC camera on the 4m Blanco telescope.

Although the median profile has lower amplitude than these three individual examples, it has a similar shape. The lower amplitude is the result of the fact that other galaxies in this luminosity range have extremely low total counts and noisy, fragmented profiles (shown by dotted lines in Fig.~\ref{fig: GC bin 7}). There are only 7 galaxies in this bin in total (due to the small volume within our limiting radius of $30\,\Mpc$), so these profiles make a significant contribution to the average. They suffer particularly badly from the background subtraction and detection effects described in the main text. This luminosity range includes massive late types in relatively low mass halos as well as early-type BCGs; the complex structure in the disks of the brightest late-types makes the systematic effects even worse\footnote{The luminosities and stellar masses of these galaxies are also somewhat harder to estimate, due to dust, complex surface brightness profiles and color gradients} than they are in the BCGs (which have higher surface brightness but simplers structure). The lower luminosity bins also include galaxies that suffer from these effects. However, with decreasing luminosity, such galaxies are increasingly outweighed in the average by those with more regular profiles.

The effect of incompleteness on the three BCG profiles is evident from the differences relative to the literature data in Fig.~\ref{fig: GC bin 7}. The central regions ($\lesssim R_{50}$) are clearly incomplete, typically with no GCs detected; around $\lesssim R_{50}$ there is a slight enhancement of counts; at $10<r<30$~kpc there appears to be a deficit of counts; and the agreement between our profiles and the literature is generally best at larger radii, particularly with regard to the slope of the profiles. These differences with the literature profiles are similar for all three BCGs, suggesting that the systematic effects responsible for them may also be similar.
\begin{figure}
    \includegraphics[width=\linewidth]{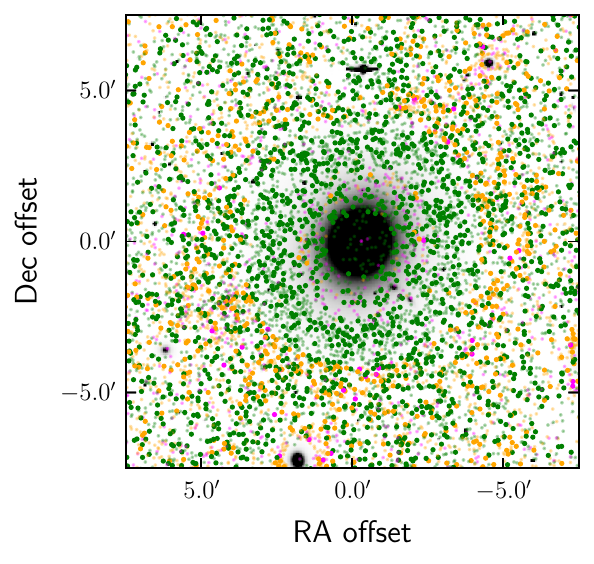}
    \caption{Image of DESI-LS sources and our selected GCs around NGC4468/M87. The sources are classified by the DESI-LS pipeline, as: point sources (PSF, green dots),  `round exponential' sources of fixed size (REX, orange dots), and extended types (exponential, de Vaucouleurs etc.; magenta dots). The selected GCs are marked with bold colors, the remaining sources are with lighter colors.}
    \label{fig:bcg_example}
\end{figure}
Fig.~\ref{fig:bcg_example} shows an image of  NGC 4486 (M87), with the full DESI-LS source catalog over-plotted to illustrate these effects. There are very few DESI-LS sources detected in the central region of the galaxy, mostly classified as point source (PSF) type by \textit{the Tractor} (the source modeling component of the DESI-LS pipeline). None of these are selected as GC candidates. Continuing outwards, there is a sparse ring of sources around $R_{50}$, which likely comprises both genuine clusters and incorrectly deblended fluctuations in the underlying galaxy light. These sources are mostly classified as PSF type at the inner edge of the ring with a greater number of extended types as the radius increases. In this region, there are also many visual surface brightness fluctuations, including compact sources, that are not associated with a source by Tractor. At larger radii, essentially all visually obvious sources appear in the catalog; in this region we see a transition to a higher-density zone of predominantly PSF sources, then a notable increase in the fraction of REX sources (`round exponential'; shown in orange), before the mix of source types finally becomes more uniform in the outskirts. These rings dominated by particular source types clearly correspond to systematic effects in the Tractor pipeline in regions of high surface brightness, and can be identified approximately with the features in Fig.~\ref{fig: GC bin 7}. In particular, the tendency for sources around $\sim5-10$~kpc ($\sim 1-2$ arcmin)
to be classified as extended likely contributes to the apparent deficit of GC candidates in this region. Although these detection effects are problematic, it seems likely that they could be substantially improved by an independent reduction of the DESI-LS images in these regions and/or adjustments to the Tractor algorithm.

\section{Detection effects around low-mass galaxies}
\label{sec:low_mass_galaxy_effects}

\begin{figure*}
  \centering

    \includegraphics[width=\linewidth]{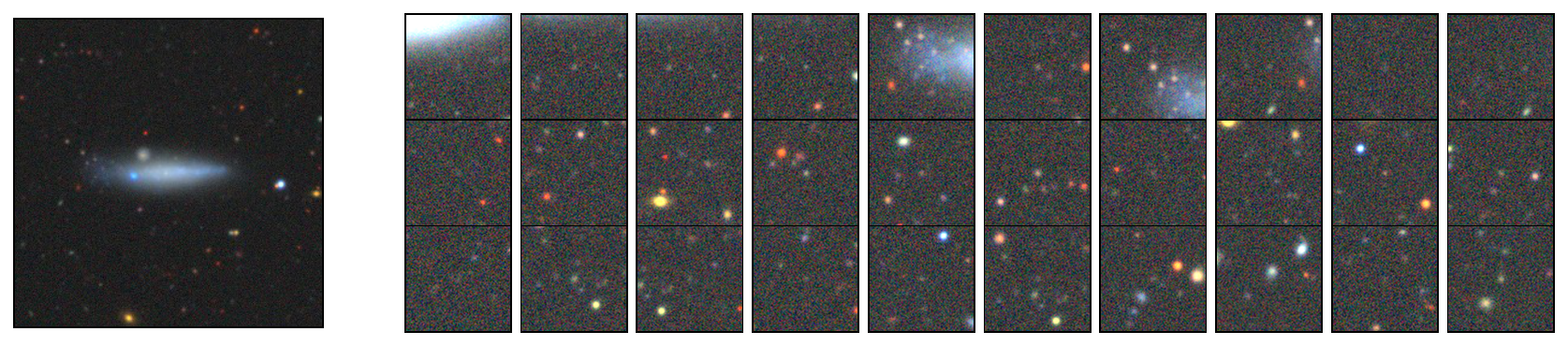}
    \vspace{0.5cm}

    \includegraphics[width=\linewidth]{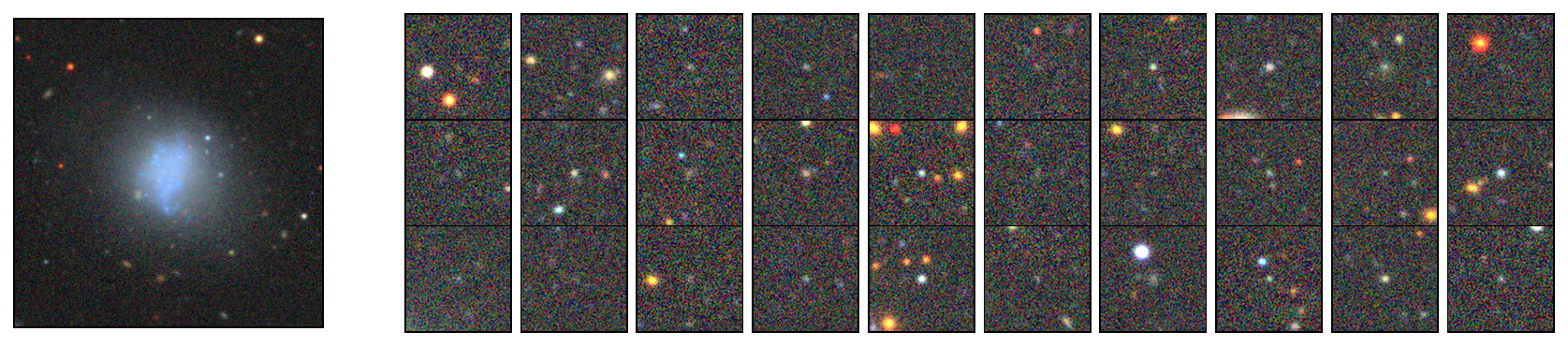}
    \vspace{0.5cm}

    \includegraphics[width=\linewidth]{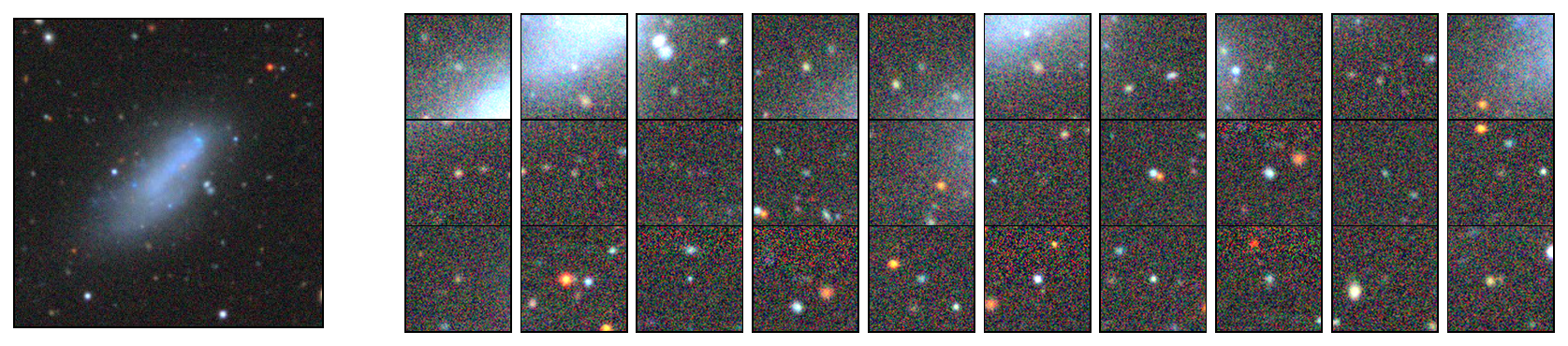}
    \vspace{0.5cm}

    \includegraphics[width=\linewidth]{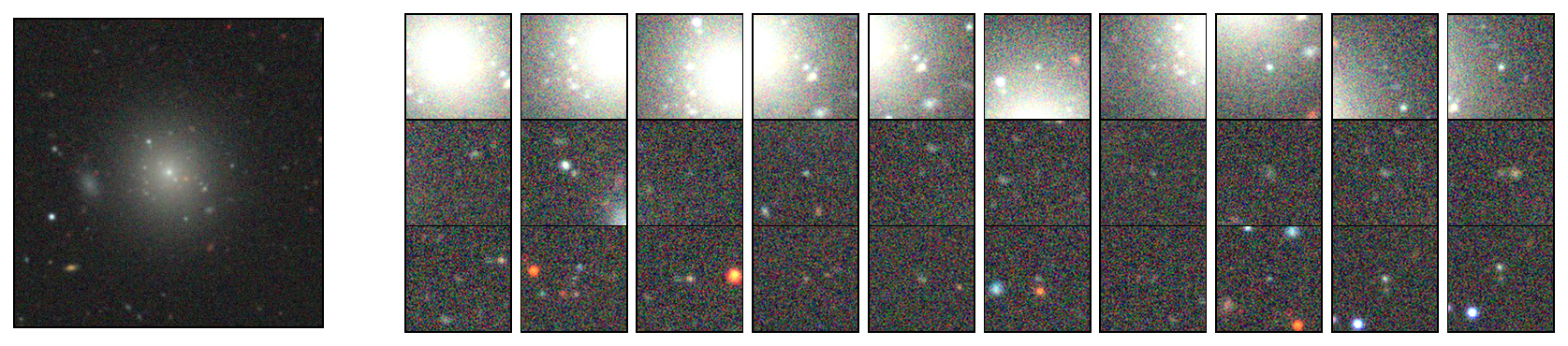}
    \vspace{0.5cm}

    \includegraphics[width=\linewidth]{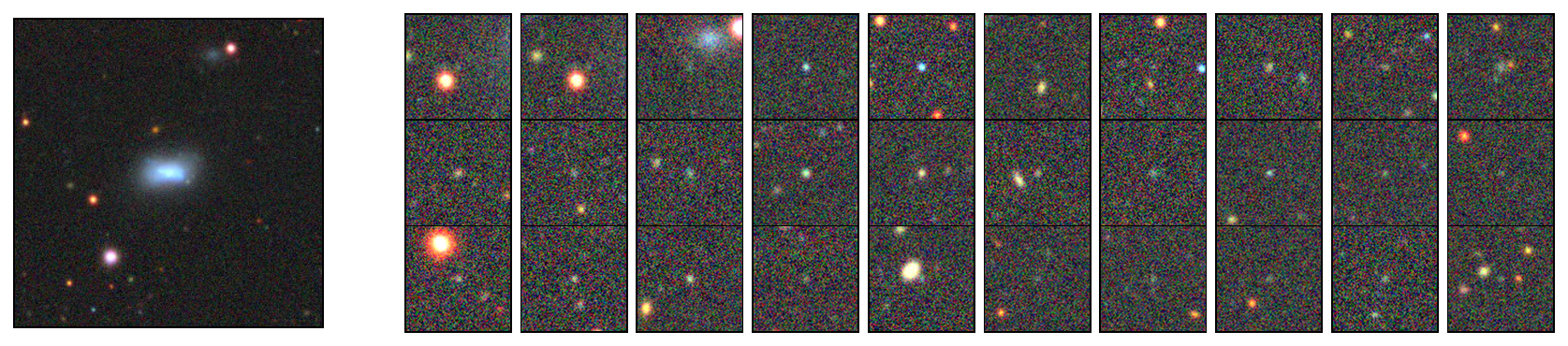}

    \caption{GC candidates around five random primaries (UGC01075, UGC05797, UGC08382, PGC041782, PGC169954) in our lowest luminosity bin. Note that most of these candidates are likely to be faint stars or galaxies; the GC counts we report in the main text are the statistical excess of sources relative to the background at large distances. PGC041782 (second from the bottom) provides an example in which GC candidates close to the primary make a significant contribution to this excess (contrast with UGC05797, second from top).
    \label{fig: GC around dwarf primaries}}
\end{figure*}

Fig.~\ref{fig: GC around dwarf primaries} shows a gallery of the 30 nearest GC candidates around 5 randomly selected galaxies in the luminosity range $8 < \log_{10} L/\lsun < 8.5$. This highlights the central star-forming regions and other surface brightness fluctuations associated with the galaxy that might (but generally do not) pass our GC candidate cuts. As discussed in the main text, sources associated with such fluctuations (both rela and spurious) are not found in the background, and hence may contribute a spurious excess of GC counts. Such sources may, potentially, explain the excess counts we find on average for low mass galaxies, relative to \citetalias{Harris_2017}, as shown in Fig.~\ref{fig: NGC-Mh}. However, in these five randomly chosen examples, only PGC041782 (a nucleated dwarf spheroidal near M86) has a significant number of sources overlapping the luminous body of the galaxy, and most of those appear to be real point-like sources.

\begin{figure}
\includegraphics[width=\linewidth]{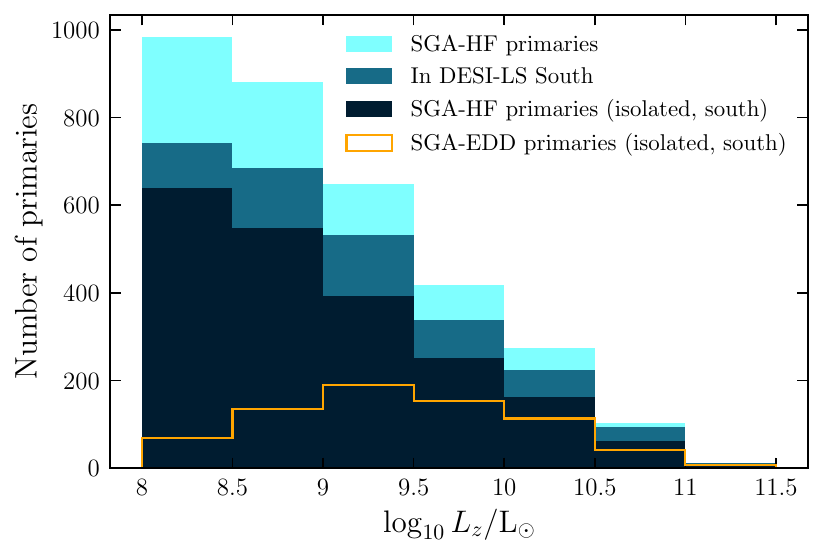}
\caption{The luminosity distribution of primary galaxies in the SGA-HF sample (black histogram), as shown for our fiducial SGA-EDD sample in Fig.~\ref{fig: primary selection}. The cyan and teal histograms correspond to supersets of the primary sample as described in the text. The orange histogram shows the distribution of SGA-EDD primaries for comparison. Removing the requirement of an EDD distance increases the sample size in the lowest luminosity bins, but does not significantly increase the sample at higher luminosity because of the restriction to distances $<30$~Mpc.  
\label{fig: primary sample no EDD}}
\end{figure}

\section{Expanded primary sample with Hubble-flow distances}
\label{sec:appendix_sample}

\begin{figure}
    \includegraphics[width=\linewidth]{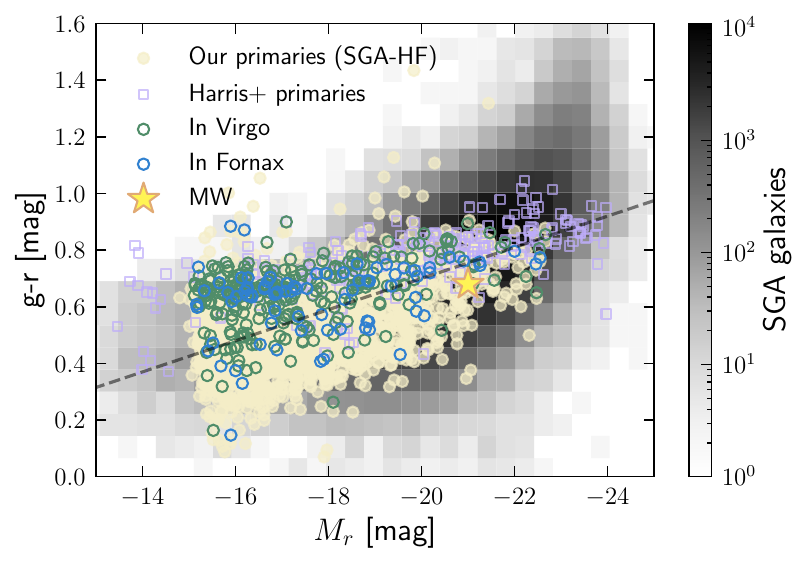}
    \caption{The color -- magnitude diagram for the SGA-HF sample (yellow circles), as shown for our fiducial SGA-EDD sample in Fig.~\ref{fig: color-mag of primaries}. The distribution of all SGA galaxies is shown in greyscale. The lines, colors and symbols are described in the caption of Fig.~\ref{fig: color-mag of primaries}.}
    \label{fig: color-mag of SGA primaries}
\end{figure}

\begin{figure}
    \includegraphics[width=\linewidth]{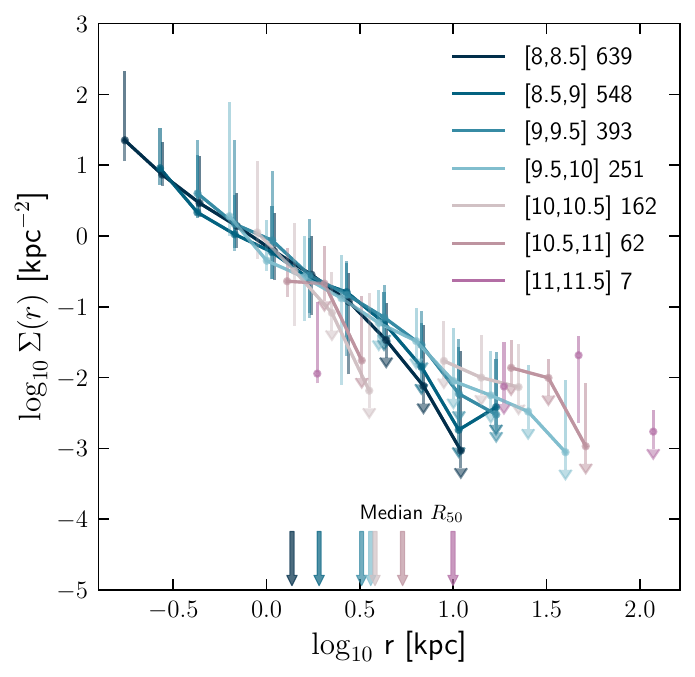}
    \caption{Stacked average density profiles for primary galaxies in the SGA-HF sample, in bins of luminosity, as show for our fiducial SGA-EDD sample in  Fig.~\ref{fig: GC profiles}.
    \label{fig: GC profile SGA sample}}
\end{figure}

\begin{figure}
    \includegraphics[width=\linewidth]{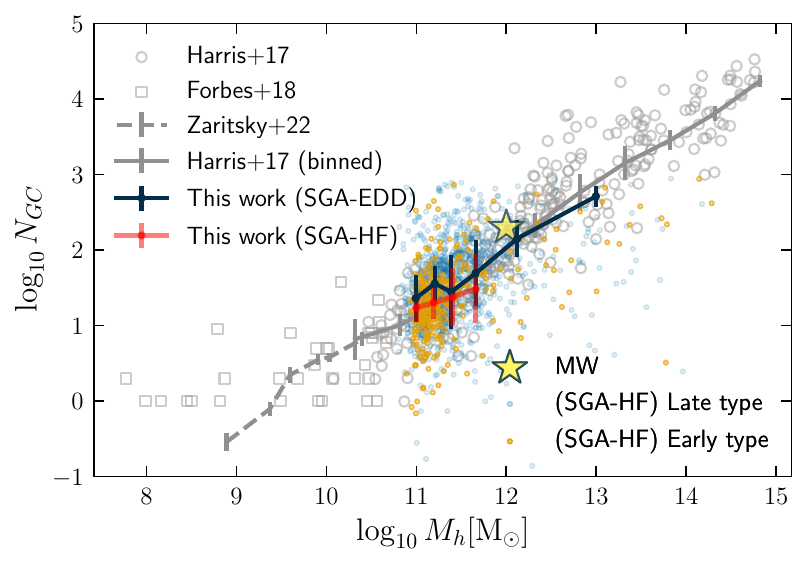}
    \caption{The \gcnumberhalomass{} relation, as Fig.~\ref{fig: NGC-Mh}, but here showing results for all isolated SGA galaxies in the area we consider, using Hubble-flow distances rather than distances from EDD. This SGA-HF sample is considerably larger, but with greater systematic uncertainties (see text.) The red line shows the median and MAD error of GC counts (derived from the stacked average profiles, not the straight average of the points) for the four lowest luminosity bins (see text).
    \label{fig: NGC-Mh relation SGA sample}}
\end{figure}

This appendix describes the expanded sample of primaries introduced in section~\ref{sec:discussion_differences_in_primary_sample}. This sample (which we call SGA-HF to distinguish it from our fiducial SGA-EDD sample) is based on Hubble flow distances, rather than redshift-independent distances. We show GC count results for this sample, for comparison to the results in section~\ref{sec:results}.

If we do not restrict our sample to galaxies with EDD distances, our sample size increases from 707 to 2062 galaxies. The luminosity distribution of the SGA-HF is shown in Fig.~\ref{fig: primary sample no EDD}; most of the additional galaxies are in the lowest luminosity bins. In the higher luminosity bins, the SGA-HF sample has roughly the same number of primaries as in the SGA-EDD sample. The color-magnitude distribution of the SGA-HF sample (Fig.~\ref{fig: color-mag of SGA primaries}) is more representative of a volume-limited population of bluer, lower-luminosity galaxies; the remaining differences with the full SGA catalog are due to our isolation criterion, as described in the main text. 

Fig.~\ref{fig: GC profile SGA sample} shows the stacked density profiles of the SGA-HF sample in each luminosity bin. Compared to Fig.~\ref{fig: GC profiles}, there is less differentiation between the outer regions of the profiles, although the systematic increase in amplitude with luminosity beyond $\sim R_{50}$ is evident. The profiles of the three most luminous bins ($\log_{10}\, L/\mathrm{L_{\odot}} > 10$) have more pronounced `gaps', and correspondingly lower integrated counts, for reasons described below.

Fig.~\ref{fig: NGC-Mh relation SGA sample} shows, in red, the median \gcnumberhalomass{} relation for this sample, with errorbars representing the MAD.  Given the greater uncertainty in the most luminous bins for these stacks, we only show this relation for the four lowest luminosity bins. Here the SGA-HF relation is consistent with our result for the SGA-EDD sample (Fig.~\ref{fig: NGC-Mh}) within $1\sigma$, but systematically lower in all bins, closer to the \citetalias{Harris_2017} average. At face-value, one interpretation of this (rather marginal) result is that, at fixed halo mass, bluer dwarf galaxies occupy dark matter halos that collapse at relatively later epochs, and swhich may be less likely to host significant GC populations. However, as discussed in the following paragraph, such an interpretation is not straightforward.

We do not show results for the higher mass bins because, with the expanded sample, these bins suffer even more from the contribution of galaxies with `catastrophic' detection effects (as described in appendix~\ref{sec:massive_galaxy_effects}. The more such cases, the less effective the $5\sigma$ trimming we apply to mitigate their effect on the average (see section~\ref{sec: density profile}). The result is much lower average counts in higher luminosity bins. Eddington bias due may also contribute to this effect: the much larger number of low-mass galaxies in the SGA-HF sample increases the probability of intrinsically GC-poor dwarfs with underestimated distances and hence overestimated stellar masses contributing to higher luminosity bins.

The scatter between Hubble flow distances and the redshift-independent distances from EDD is sufficiently large that many galaxies in the SGA-EDD sample are redistributed to different luminosity bins in the SGA-HF sample. This effect is asymmetric because we impose a cut in distance at 30~Mpc, such that galaxies for which the HF distance is an over-estimate drop out of the sample. At fixed apparent magnitude, galaxies for which the HF distance is an under-estimate are assigned lower luminosities, hence lower masses, and therefore, ultimately, lower virial masses. A lower distance estimate also implies a smaller GCLF correction for a given primary, because the faintest cluster candidate is assigned a relatively lower absolute magnitude. Thus distance underestimates lead to lower $N_{GC}$. 

Eddington bias may also have an effect at lower luminosities in the SGA-EDD sample. There is a large reservoir of intrinsically faint systems with $\log_{10} L < 7.5$ and few (or zero GCs). Those with over-estimated distances may reduce the average counts in the lowest luminosity bins. Since there are no very faint systems with EDD distances, the EDD catalog does not suffer from this (potential) bias. 

More thorough forward-modeling of the sample will be required to disentangle these effects. Since that modeling is beyond the scope of our work here, we have chosen focus on from the SGA-EDD sample in the main text, even though it may be less representative of the  galaxy population overall.

\bibliography{refs}{}
\bibliographystyle{aasjournal}



\end{document}